\documentclass[pdftex,twocolumn,epjc3]{svjour3}          

\RequirePackage[T1]{fontenc}

\smartqed  

\newcommand{\G}{{\cal G}}
\usepackage{amssymb}
\usepackage{float}
\usepackage{amsmath}
\usepackage{setspace}
\usepackage{verbatim}
\usepackage{graphicx}
\usepackage{multirow}
\usepackage{rotating}
\usepackage{booktabs}
\usepackage{chngpage}
\RequirePackage{graphicx,subfigure}
\usepackage{graphicx}

\newcommand{\bef}{\begin{figure}}
\newcommand{\eef}{\end{figure}}
\newcommand{\bec}{\begin{center}}
\newcommand{\eec}{\end{center}}
\newcommand{\be}{\begin{equation}}
\newcommand{\ee}{\end{equation}}
\newcommand{\rf}[1]{(\ref{eq:#1})}
\usepackage{rotating,graphicx}
\usepackage{dcolumn}
\usepackage{bm}
\usepackage{pbox}

\usepackage{bigstrut} 
\RequirePackage{graphicx}
\RequirePackage{mathptmx}      
\RequirePackage{flushend}
\RequirePackage[numbers,sort&compress]{natbib}
\RequirePackage[colorlinks,citecolor=blue,urlcolor=blue,linkcolor=blue]{hyperref}
\usepackage{graphicx}
\usepackage{multirow}
\usepackage{booktabs}
\usepackage{chngpage}
\RequirePackage{graphicx,subfigure}
\usepackage{graphicx}
\usepackage{hyperref}

\begin{document}

\title{Linear Nash-Greene fluctuations on the evolution of $S_8$ and $H_0$ tensions}

\author{Abra\~{a}o J. S. Capistrano\thanksref{e1,addr1,addr2}
        \and
       Luís A. Cabral\thanksref{e2,addr3,addr2}
       \and
       Jos\'{e} A. P. F. Mar\~{a}o\thanksref{e3,addr4}
       \and
        Carlos H. Coimbra-Ara\'{u}jo\thanksref{e4,addr1,addr2}
}

\thankstext{e1}{e-mail: capistrano@ufpr.br}
\thankstext{e2}{e-mail: cabral@uft.edu.br}
\thankstext{e3}{e-mail: jose.marao@ufma.br}
\thankstext{e4}{e-mail: carlos.coimbra@ufpr.br}

\institute{Universidade Federal do Paran\'{a}, 85950-000, Palotina-PR, Brazil.\label{addr1}
                    \and
          Applied physics graduation program, UNILA, 85867-670, Foz do Igua\c{c}u-PR, Brazil\label{addr2}
                      \and
          Curso de Física, Setor Cimba, Universidade Federal do Tocantins, 77824-838,  Araguaína-TO, Brazil \label{addr3}
                       \and
          Centro Tecnológico, Departamento de Matemática, Universidade Federal do Maranhão, 65085-580, São Luís-MA, Brazil\label{addr4}
}

\date{Keywords: Nash-Greene embeddings, gravitation, Dark energy}

\maketitle

\begin{abstract}
We present the perturbation equations in an embedded four space-time from the linear Nash-Greene fluctuations of background metric. In the context of a five-dimen-\\sional bulk,  we show that the cosmological perturbations are only propagated by the gravitational tensorial field equation. In Newtonian conformal gauge, we study the matter density evolution in sub-horizon regime and on how such scale may be affected by the extrinsic curvature. We apply a joint likelihood analysis to the data by means of the Markov Chain Monte Carlo (MCMC) method for parameter estimation using a pack of recent datasets as the Pantheon Supernovae type Ia, the Baryon Acoustic Oscillations (BAO) from DR12 galaxy sample and Dark Energy Survey (DESY$1$). We discuss the tensions on the Hubble constant $H_0$ and the growth amplitude factor $S_8$ of the observations from Planck 2018 Cosmic Microwave Background (CMB) and the local measurements of $H_0$ with Hubble Space Telescope (HST) photometry and Gaia EDR3.  As a result, we obtain an alleviation below $\sim 1\sigma$ in the contours ($S_8$-$\Omega_m$) at $68.4\%$ confidence level (CL) when compared with DESY1 data. On the other hand, the $H_0$ tension persists with $\sim 2.6\sigma$ at $68.4\%$ CL and $1.95\sigma$ at $95.7\%$ CL, aggravated with the inclusion of BAO data.
\end{abstract}

\section{Introduction}
The Occam's razor is one of the cornerstone philosophical principles in science that states that the simplest solution of a problem should be adopted in detriment of  complex ones. In this realization, the $\Lambda$CDM model has been the simplest and successful solution to deal with the accelerated expansion of the universe as corroborated for several independent observations in the last two decades \cite{sahni,will,alamet,kowal,jaff,izzo,george,allen,bax,ricardo,planck2018}. Despite its success, the $\Lambda$CDM model has important drawbacks that must be taken into account. For instance, the main components of $\Lambda$CDM model lack a fundamental explanation about the nature of the cosmological constant $\Lambda$ and also the cold Dark Matter (CDM) problem \cite{nemiroff, santos, kumar, velten, sultana, Siva, Nozari}.

In this paper, we propose a model that adds a new curvature to the standard Einstein gravity by means of local dynamical embeddings. As a first test, we focus on studying  the problem of the appearance of tensions at several standard deviations ($\sigma$) of the growth amplitude factor $S_8=\sigma_8\left(\Omega_{m0}/0.3\right)^{0.5}$ and the Hubble constant $H_0$ revealed by the mismatch of the data inferred from Cosmic Microwave Background (CMB) radiation probe and the large scale structure (LSS) observations, considering the concordance $\Lambda$CDM model as a background. The $\sigma_8$ is the r.m.s amplitude of matter density at a scale of a radius $R\sim 8$h.Mpc$^{-1}$ within a enclosed mass of a sphere \cite{fan} and $\Omega_m$ denotes the matter density cosmological parameter. The main problem apparently resides in the fuzzy origin of such mismatch, which could be a result from systematics or due to deviations of gravity. In addition, we also analyse the consequences on the Hubble tension that goes from 4-$\sigma$ to 6-$\sigma$ standard deviations of statistical distance between local measurements of the Hubble constant $H_0$ and CMB Planck data (see ref.\cite{valentino} for review and references thereof). Such impasse still resists in both early and late universe landscapes even if one does not consider the Planck CMB data \cite{addison} and evidences that similar discrepancies may also occur in the matter distributions around 2-$\sigma$ \cite{planck2018,battye, birrer, mccarthy} between the growth amplitude factor $\sigma_8$ and the matter content $\Omega_m$. Moreover, the structure growth parameter $S_8$ also presents large discrepancies as measured by Planck probe compared with surveys as KiDS 450 \cite{KiDs1,KiDs2,KiDs3}, DESY1\cite{DES,DES2} and CFTLens \cite{CFHT1,CFHT2,CFHT3}. In the recent KiDS 1000\cite{KiDs1000}, this discrepancy persists around $3\sigma$.  In particular, to avoid a biased dependence of $\sigma_8$, the quantity $f\sigma_8(z)$ is a good model-independent discriminator for mapping the growth rate of matter. This alleged discrepancy opens an interesting arena for testing gravitational models, once the possibility to alleviate such tensions may come from modified gravity and/or their extensions \cite{Eva,Nesseris2017,kazan,radou,divalentino,lambiase,sebastian,ikeda,konitopoulos}.

In the context that gravity may be modified departed from Einstein gravity or other fundamental principle, we explore the embedding of geometries (or hypersurfaces) to elaborate a model independent based on seminal works on the subject \cite{maia2, GDE, QBW} in order to tackle the aforementioned tensions in the problem of explanation of the accelerated expansion of the universe. In hindsight, the seminal problem of embedding theories lies in the hierarchy problem of fundamental interactions. The possibility that gravity may access extra-dimensions is taken as a principle for solving the  huge  ratio  of  the  Planck masses ($M_{Pl}$) to  the  electroweak  energy  scale $M_{EW}$ in such $M_{Pl}/M_{EW}\sim 10^{16}$. This option has been explored more vigorously in the last two decades as a candidate for solution of the dark energy paradigm. Most of these models have been Kaluza-Klein and/or string inspired, such as, for instance, the works of the Arkani-Hamed, Dvali and Dimopolous (ADD) model \cite{add}, the Randall-Sundrum model \cite{RS,RS1} and the Dvali-Gabadadze-Porrati model (DPG) \cite{dgp}. Differently from these models with specific conditions, and apart from the braneworld standards and variants, we have explored the embedding as a fundamental guidance for elaboration of a gravitational physical model. Until then, several authors explored the embedding of geometries and its physical consequences as a mathematical structure to apply to gravitational problems \cite{Brandon, maia2, GDE, QBW, sepangi, sepangi1, maiabook, gde2, sepangi2, cap2014, jalal2015, capistrano2015, capistrano2016a, capistrano2016b, cap2017, capistrano2017,capistrano2019}.

The plan of the paper is organised in sections. In the second section, we revise the embedding of geometries and on how it may be used to construct a physical framework. In this context, the Nash-Greene theorem is discussed. The third and fourth sections verse on the background Friedmann\\–Le-ma\^{\i}tre–Robertson–Walker (FLRW) metric, transformations and gauge variables also developed involving the extrinsic curvature, respectively. The fifth section shows the resulting conformal Newtonian gauge equations. In the sixth section, we show the contrast matter density $\delta_m(a)$ as a result of the Nash fluctuations and an effective Newtonian constant $G_{eff}$ is also determined that carries a signature of the extrinsic curvature. In the seventh section, we analyze both Hubble and $S_8$ tensions. By means of a MCMC sampler relying on the Code for bayesian analysis (\texttt{Cobaya}\footnote{\url{https://github.com/CobayaSampler/cobaya}}) \cite{cobaya1, cobaya2} to constrain the posteriors, we make a joint analysis of data taking into account the evolution of background parameter $H(z)$ and $\Omega_m$ distributions. From CMB fluctuations, we adopted Planck 2018 data \cite{planck2018} and consider high+low multipoles from CMB temperature and polarisation angular power spectra, i.e., high-\emph{l}.plik.TTTEEE, low-\emph{l} EE polarisation and low-\emph{l} TT temperature. To compute growth data, $H(z)$ and BAO, we further consider SNIa Pantheon data \cite{Pantheon}, clustering and weak lensing from DESY1 \cite{DES} and the DR12 ``consensus'' galaxy sample \cite{DR12}. As local measurements on $H_0$, we adopt Riess et al. 2020 \cite{riess20} data from Hubble Space Telescope (HST) photometry and Gaia EDR3 parallaxes. The MCMC chains are analysed by using \texttt{GetDist}\footnote{\url{https://github.com/cmbant/getdist}} \cite{getdist}.  As criteria for model selection, we use three information criteria such as the Akaike Information Criterion (AIC) \cite{Akaike}, Modified Bayesian Information Criteria (MBIC) \cite{bic, mbic} and Hannan-Quinn Criterion (HQC) \cite{hqc}. In the final section, we present our remarks and prospects.

It is noteworthy to point out that we adopt the Landau spacelike convention $(-+++)$ for the signature of the four dimensional embedded metric and speed of light $c=1$. Concerning notation, capital Latin indices run from 1 to 5. Small case Latin indices refer to the only one extra dimension considered. All Greek indices refer to the embedded space-time counting from 1 to 4. Hereon we indicate the non-perturbed (background) quantities by the upper-script symbol ``0''.

\section{The induced embedded equations}
In the following subsections, we present our theoretical framework. First, the embedding of geometries is presented and institutes a mathematical background landscape. Secondly, by pursuing this intent, the induced field equations of the embedded space-time are presented that result from the integrability of the embedding given by Nash-Greene theorem to arise a viable physical framework.
\subsection{The Einstein-Hilbert principle for a five dimensional bulk}
Although embeddings can be made in an arbitrary number of dimensions (see \cite{maia2,GDE,QBW,gde2,maiabook,jalal2015,capistrano2015,capistrano2016a,capistrano2016b,capistrano2017,capistrano2019}), the current alternative models of gravitation are normally stated in five dimensions with one degree of freedom. Then, we start with a model defined by a gravitational action S in the presence of confined matter field of a four-dimensional space-time embedded in a five-dimensional larger space as
\begin{equation}\label{eq:action}
S= -\frac{1}{2\kappa^2_5} \int \sqrt{|\mathcal{G}|}^5\mathcal{R}d^{5}x - \int \sqrt{|\mathcal{G}|}\mathcal{L}^{*}_{m}d^{5}x\;,
\end{equation}
where $\kappa^2_5$ is a fundamental energy scale on the embedded space, $^5\mathcal{R}$ denotes the five dimensional Ricci scalar of the bulk and $\mathcal{L}^{*}_{m}$ denotes the confined matter Lagrangian in such the matter energy momentum tensor fulfills a finite hypervolume with constant radius $l$ along the fifth-dimension. In the first term in Eq.(\ref{eq:action}), $^5\mathcal{R}$ can be expressed in terms of the intrinsic and extrinsic geometric quantities, and the action S is rewritten as
\begin{equation}\label{eq:action2}
S= -\frac{1}{2\kappa^2_5} \int \sqrt{|\mathcal{G}|}(R-K^2+h^2)d^{5}x - \int \sqrt{|\mathcal{G}|}\mathcal{L}^{*}_{m}d^{5}x\;,
\end{equation}
with $R$ is the four-dimensional Ricci scalar, and the extrinsic quantities as $K^{2}=k^{\mu\nu}k_{\mu\nu}$ is the Gaussian curvature and the mean curvature $h^2=h\! \cdot \! h$ and $h= \;g^{\mu\nu}\;k_{\mu\nu}$, where $g_{\mu\nu}$ is the four-dimensional metric and $k_{\mu\nu}$ is the extrinsic curvature. It is important to note that the form of the action in Eq.(\ref{eq:action2}) results from a general process of embedding of geometries, as shown in detail in references \cite{maia2,GDE,QBW,gde2} for the embedding of a four-dimensional space-time into a D-dimensional space-time . To obtain the field equations,  the variation of Einstein-Hilbert action in Eq.(\ref{eq:action}) with respect to the bulk metric $\mathcal{G}_{AB}$ leads to the five-dimensional Einstein equations
\begin{equation}\label{eq:EEbulk}
^5\mathcal{R}_{AB} - \frac{1}{2}\;^5\mathcal{R} \mathcal{G}_{AB}= \alpha^{\star}\mathcal{T}_{AB}\;,
\end{equation}
where $\alpha^{\star}$ is the energy scale parameter and $\mathcal{T}_{AB}$ is the energy-momentum tensor for the bulk \cite{GDE,gde2,QBW,maia2}, and the bulk metric $\mathcal{G}_{AB}$ is assumed as
\begin{eqnarray}
  \mathcal{G}_{AB} &=& \left(
          \begin{array}{cc}
            g_{\mu\nu} & 0 \\
            0 & 1 \\
          \end{array}
        \right)\;,
   \label{eq:metricbulk}
\end{eqnarray}
with $g_{55}=1$ and the extra-dimensional indices are fixed to 1, since in this application we have only one extra-dimension. In accordance with the Nash-Greene theorem \cite{Nash,Greene}, orthogonal perturbations of the metric induce the appearance of the extrinsic curvature in that direction. To our purposes, we are restricted to the fourth dimensionality of the space-time embedded in a five dimensional bulk space following the confinement hypothesis \cite{Donaldson,Taubes}. Such dimensionality will suffice based on experimentally high-energy tests \cite{lim}.

This  model  can be regarded as a four dimensional hypersurface dynamically evolving  in  a five-dimensional bulk with constant
curvature whose related Riemann tensor is
\begin{equation}\label{eq:bulkconst}
^5\mathcal{R}_{ABCD}=
K_*\left(\mathcal{G}_{AC}\mathcal{G}_{BD}-\mathcal{G}_{AD}\mathcal{G}_{BC}\right),
\;\;\;A...D =  1...5\;,
\end{equation}
where $\mathcal{G}_{AB}$ denotes the bulk metric components in arbitrary coordinates and the  constant  curvature $K_{\ast}$ is
either zero (flat bulk) or it can have   positive (deSitter) or negative (anti-deSitter) constant curvatures. In accordance with recent observations \cite{planck2018}, the cosmological constant $\Lambda$ has a very small value but we do not consider any dynamical contribution from it. The permanence of $\Lambda$ is just for completeness purposes and will be omitted henceforth in the induced four-dimensional field equations. The  bulk geometry is  actually  defined by the Einstein-Hilbert principle in Eq.(\ref{eq:action}), which  leads  to Einstein's equations for the bulk as shown in Eq.(\ref{eq:EEbulk}).

In this sense, it is possible to search a more general physical theory based on the geometries of embedding. Although it is not explicitly showed here, depending on the type of the embedding (e.g., local or global, isometric, analytic or differentiable, etc.), braneworld models may be an example of this framework \cite{GDE}. Another important aspect of the original Nash embedding is that it is applied to a flat $D$-dimensional Euclidean space. It was explored in a work by J. Rosen \cite{jrosen} with an analysis on pseudo-Euclidean spaces. From its generalisation of pseudo-Riemannian manifolds to non-positive signatures results that the embedding of the space-times  may  need  a  larger  number  of  dimensions, which was made only two decades later by Greene \cite{Greene}. Hereon, we simply call the Nash-Greene theorem.

In a nutshell, the  smoothness of the embedding is the cornerstone concept of the Nash-Greene theorem, once this embedding results from a differentiable mapping of functions of the manifolds. On the other hand, it is not capable of telling us about the physical dynamic equations or evolution of the gravitational field by its own. Thus, a natural  choice  for  the  bulk is that its  metric satisfies the Einstein-Hilbert  principle. By design, it represents the variation  of the Ricci scalar and the related curvature must be ``smoother'' as possible \cite{maiabook}. It warrants that the embedded geometry and their deformations  will also be differentiable. To obtain the induced field equations, we firstly need to calculate the tangent components of Eq.(\ref{eq:EEbulk}). To do so, the embedding process must be properly defined as we show in the following.

\subsection{The integrability of the embedding}
Let a  Riemannian  manifold  $V_4$ be endowed with a non-pertur-bed metric $^{(0)}g_{\mu\nu}$ being locally and isometrically  embedded in a  five-dimensional  Riemannian  space $V_5$. Given a  differentiable  and  regular map $\mathcal{X}: V_4 \rightarrow  V_5$, one imposes the  embedding  equations
\begin{eqnarray}
&&\mathcal{X}^{A}_{,\alpha} \mathcal{X}^{B}_{,\beta}\mathcal{G}_{AB} = g^{(0)}_{\alpha\beta}\;, \label{eq:imersao 1}\\
&&\mathcal{X}^{A}_{,\alpha}\;^0\eta^{B}_{a}\mathcal{G}_{AB} = 0\;, \label{eq:imersao 2} \\
&&^0\eta^{A}_{a}\;^0\eta^{B}_{b}\mathcal{G}_{AB}=1\;,\label{eq:imersao 3}
\end{eqnarray}
where we have  denoted $\mathcal{X}^{A}$ the non-perturbed embedding coordinate, $\G_{AB}$  the metric components of  $V_5$  in  arbitrary  coordinates, and $^0\eta$  denotes  the non-perturbed unit  vector field orthogonal  to $V_4$. This mechanism avoids possible coordinate gauges that may drive to false perturbations. The colon signs denote ordinary derivatives.

The meaning of those former set of equations is that Eq.(\ref{eq:imersao 1}) represents the isometry condition between the bulk and the embedded space-time. The orthogonality between the embedding coordinates $\mathcal{X}$ and $^0\eta$ is represented in Eq.(\ref{eq:imersao 2}). Moreover, Eq.(\ref{eq:imersao 3}) denotes the set of vectors normalisation $^0\eta$. As a result, the  integration of the set of  Eqs. (\ref{eq:imersao 1}), (\ref{eq:imersao 2}) and (\ref{eq:imersao 3}) gives  the embedding  map $\mathcal{X}$.

The relation between the geometries involved in (dynamical) embedding naturally leads to the appearance of new geometric objects. Based on in traditional textbooks \cite{eisen}, one of the fundamental objects is the extrinsic curvature. The extrinsic  curvature of the embedded space-time $V_4$ is the projection of  the  variation of the vector $^0\eta$ onto the tangent plane such as
\be
k^{(0)}_{\mu\nu} =  -\mathcal{X}^A_{,\mu}\;^0\eta^B_{,\nu} \G_{AB}= \mathcal{X}^A_{,\mu\nu}\;^0\eta^B \G_{AB}\;. \label{eq:extrinsic}
\ee
The main concern of our work is provide a complement to the Einstein gravity by adding the  extrinsic  curvature and study its implications to a physical theory in five dimensions.

\subsubsection{The dynamical embedding: the Nash flow}
The dynamical embedding mainly reflects on how the ambient space (the bulk) is related to the embedded space-time. In this work, we consider a five-dimensional bulk with constant as in Eq.(\ref{eq:bulkconst}) with an evolving embedded four space-time as summarised in the following geometrical process. The concept of ``pure'' Nash deformations is that they are gauge-free since they access the ambient space and are generated by perturbations along the direction orthogonal to $V_4$ to filter out any coordinate gauges. In five dimensions, this process is simplified and just one deformation parameter (one degree of freedom) suffices to locally deform the embedded background, which can be done by the Lie transport. In a geometrical sense, this is motivated by the notion of the curvature radii of the embedded background. The single curvature radius $y_0$ is then the smallest of these solutions, corresponding to the direction in which the embedded space-time deviates  more sharply from the tangent plane. The curvature radii of the background must satisfy the homogeneous equation
\begin{equation}\label{eq:curvradii}
\mbox{det}(g_{\mu\nu}- y\; k_{\mu\nu})=0\;.
\end{equation}
This is a local invariant property of the embedded space-time and does not depend on the chosen Gaussian system \cite{eisen}. We reinforce the idea that this process occurs at the linkage of the ambient space with the deformations of the space-time and the physical consequences we will analyse after the induced quantities into the perturbed embedded (physical) space-time, where the cosmological perturbation theory applies.

As it happens, let a  geometric  object $\bar{\Omega}$ be constructed in $V_4$ in any  direction  $^0\eta$  by  the  Lie transport along the flow for a certain small distance $\delta y$. It is worth noting that it is irrelevant if the distance $\delta y$ is time-like or not, nor it is  positive or  negative. Then, the Lie transport is given by $\Omega   = \bar{\Omega}  + \delta y  \pounds_{^0\eta}\bar{\Omega}$,  where $\pounds_{^0\eta}$  denotes the Lie  derivative  with respect  to the normal vector $^0\eta$. In this sense,  the Lie  transport of the Gaussian coordinates vielbein $\{\mathcal{X}^A_\mu , ^0\eta^A \}$, defined on  $V_4$, can be written as
\begin{eqnarray}
&&\mathcal{Z}^{A}_{,\mu} =  \mathcal{X}^{A}_{,\mu}  +   \delta y \;\pounds_{^0\eta}\mathcal{X}^{A}_{,\mu}=  \mathcal{X}^{A}_{,\mu} + \delta y \;  ^0\eta^{A}_{,\mu}\;,\label{eq:pertu1}\\
&&\eta^A =\;^0\eta^A  +   \delta y\;[^0\eta, ^0\eta]^A= \;^0\eta^A \;.\label{eq:pertu2}
\end{eqnarray}
Interestingly, from  Eq.(\ref{eq:pertu2}), it is straightforward the derivative of $^0\eta$ is affected by perturbations in a sense $\eta_{,\mu} \neq\;^0\eta_{,\mu}$.

Concerning perturbations of the embedded space-time $V_4$, there is a set of perturbed coordinates $\mathcal{Z}^A$ to satisfy the embedding equations likewise Eqs.(\ref{eq:imersao 1}), (\ref{eq:imersao 2}) and (\ref{eq:imersao 3}), as
\be
\mathcal{Z}^{A}_{,\mu} \mathcal{Z}^{B}_{,\nu}\G_{AB}=g_{\mu\nu},\;  \mathcal{Z}^{A}_{,\mu}\eta^B \G_{AB}=0,  \;  \eta^A \eta^B \G_{AB}=1\;.\label{eq:Z}
\ee
As seen in the non-perturbed case, the perturbed coordinate $\mathcal{Z}$ defines a coordinate chart between the bulk and the embedded space-time.

Replacing  Eqs.(\ref{eq:pertu1}) and (\ref{eq:pertu2}) in  Eqs.(\ref{eq:Z}) and (\ref{eq:extrinsic}), for instance, we obtain the fundamentals objects of  the new  manifold in linear perturbation
\begin{eqnarray}
&&g_{\mu\nu} =  g^{(0)}_{\mu\nu}+\delta g_{\mu\nu}+...=g^{(0)}_{\mu\nu}-2 \delta y\;k^{(0)}_{\mu\nu} + ...\label{eq:g}\;,\\
&&k_{\mu\nu}  = k^{(0)}_{\mu\nu}+\delta k_{\mu\nu}+...=k^{(0)}_{\mu\nu}-2 \delta y\;^0g^{\rho\sigma}k^{(0)}_{\mu\rho}k^{(0)}_{\nu\sigma}+ ...  \label{eq:k1}
\end{eqnarray}
Hence, taking the derivative of Eq.(\ref{eq:g}) with  respect  to deformation parameter $y$ and compare with Eq.(\ref{eq:k1}), Nash flow is written as
\begin{equation}\label{eq:nashdeformation}
 k_{\mu\nu}=-\frac{1}{2}\frac{\partial g_{\mu\nu} }{\partial y}\;.
\end{equation}
This former expression can be generalized to arbitrary number of dimensions with a set of arbitrary family of orthogonal deformations $\delta y$.

\begin{figure}[htp]
\centering
\setlength{\fboxsep}{0.5pt}%
\setlength{\fboxrule}{0.5pt}%
\fbox{\includegraphics[width=3.2in, height=1.8in]{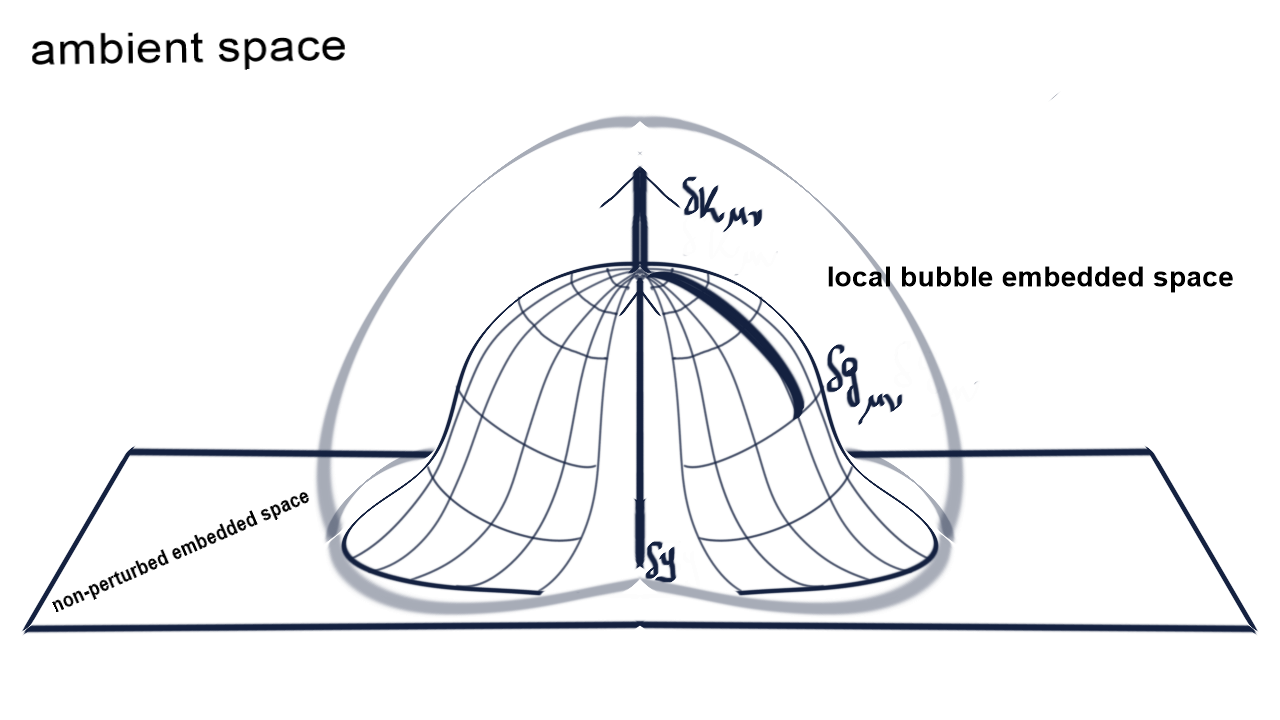}}\quad
\caption{A pictorial view of dynamical embedding in an ambient space. A ``pith helmet'' shaped-like surface emphasizes the embedding bubble formed in the local embedded space-time deformed by the perturbations of the $y$ parameter. The tangent space is orthogonally deformed and the related cosmological perturbations can be studied with a chosen gauge in $\delta g_{\mu\nu}$.}\label{fig:nash}
\end{figure}

It is noteworthy to point out that the ADM formulation gives a similar expression later discovered  by Choquet-Bruhat and J. York \cite{bruhat}. In a physical context, the interpretation of Eq.(\ref{eq:nashdeformation}) reinforces the confinement of matter as a consequence of the  well established  experimental  structure of  special  relativity,   particle  physics and  quantum  field theory,  using  only the observable which interact with the standard gauge  fields and their  dual properties. It imposes a geometric constraint that localizes the matter in $V_4$ \cite{maia2,GDE}.
It is important to note that the Nash-Greene fluctuations on the perturbed metric $g_{\mu\nu} = g^{(0)}_{\mu\nu}+\delta g_{\mu\nu}+\delta^2 g_{\mu\nu}+....$ are continuously smooth and naturally go on adding small increments $\delta g_{\mu\nu}$ to the background metric. A pictorial view of this process can be seen in Figure (\ref{fig:nash}), where the deformed embedded space-time generates a local bubble of deformation on the original background.
Unlike those of processes of rigid embedding models \cite{RS,RS1}, where the deformation parameter $y$ is commonly inserted in the line element to obtain cosmological perturbations with additional assumptions; in dynamical embeddings, the orthogonal deformations parameter $y$ accesses the ambient space and does not appear in the line element. If one insists, it will end up to inconsistencies in the perturbed embedded space-time with gauge-fixing problems.

As it happens, the resulting perturbed geometry $g_{\mu\nu}$ can be bent and/or stretch without ripping the embedded space-time, which it is not possible to do in the context of the Riemannian geometry as acknowledged by Riemann himself \cite{riemann}. This feature is exclusive to dynamical embeddings. Due to the confinement, the standard cosmological theory cannot be applied beyond the embedded space-time. More specifically, it applies to the metric perturbation $\delta g_{\mu\nu}$ and to the induced equations onto the embedded space-time. Then, we calculate the linear perturbations for five-dimensions that the new geometry $\tilde{g}_{\mu\nu}=\;g^{(0)}_{\mu\nu}+\delta g_{\mu\nu}$ generated by Nash's fluctuations is given by
\begin{equation}\label{eq:metricperturbada}
\tilde{g}_{\mu\nu}=\;g^{(0)}_{\mu\nu} - 2\delta y k^{(0)}_{\mu\nu}\;,
\end{equation}
and the related perturbed extrinsic curvature
\begin{equation}\label{curvextrperturbada}
\tilde{k}_{\mu\nu}=\;k^{(0)}_{\mu\nu}-2 \delta y\;^{(0)}g^{\sigma\rho}\;k^{(0)}_{\mu\sigma}k^{(0)}_{\nu\rho}\;,
\end{equation}
where we can identify $\delta k_{\mu\nu}=-2\delta y\;^{(0)}g^{\sigma\rho}\;k^{(0)}_{\mu\sigma}k^{(0)}_{\nu\rho}$ valid in the ambient space. Using the Nash relation $\delta g_{\mu\nu}= -2 k^{(0)}_{\mu\nu}\delta y$, $\delta y$ is replaced and we obtain
\begin{equation}\label{perturbkmunu}
\delta{k}_{\mu\nu}=\;^{(0)}g^{\sigma\rho}\;k^{(0)}_{\mu\sigma}\;\delta g_{\nu\rho}\;.
\end{equation}
This is an important result since it correctly shows how the effects of the extrinsic quantities (e.g., $\delta y$ and $\delta k_{\mu\nu}$) can be projected onto the perturbed four embedded space-time. In addition, by means of the induced field equations, as will be shown in the following, the resulting physics of the embedded space-time can be consistently studied.

\subsubsection{Integrability  conditions and induced four dimensional equations}
The  integrability  conditions of the embedding are given by the  non-trivial  components of  the  Riemann  tensor of  the embedding  space as
\begin{eqnarray}
&&^5{\cal R}_{ABCD}\mathcal{Z}^A_{,\alpha}\mathcal{Z}^B_{,\beta}\mathcal{Z}^C_{,\gamma}\mathcal{Z}^D_{,\delta}= R_{\alpha\beta\gamma\delta}+ 2k_{\alpha[\gamma}k_{\delta]\beta},\label{eq:G1}\\
&&^5{\cal R}_{ABCD}\mathcal{Z}^A_{,\alpha} \mathcal{Z}^B_{,\beta}\mathcal{Z}^C_{,\gamma}\eta^D=\;k_{\alpha[\beta;\gamma]} \;, \label{eq:C1}
\end{eqnarray}
where $^5{\cal R}_{ABCD}$ is the five-dimensional Riemann tensor. The semicolon denotes covariant derivative with respect to the metric. The brackets apply the covariant derivatives to the adjoining indices only.

The  first  equation is called Gauss equation that shows that Riemann   curvature of  bulk space  acts  as  a  reference  for  the  Riemann  curvature  of  the  embedded  space-time. The  second  equation (Codazzi equation) evinces the projection of  the  Riemann  tensor  of  the embedding space  along the  normal  direction that is  given  by  the tangent variation of  the  extrinsic  curvature. This guarantees to reconstruct the five-dimensional geometry and understand their properties from the dynamics of the four-dimensional space-time $V_4$. These equations provide the necessary and sufficient conditions for the existence of the embedded manifold. As a solution of these equations, the analyticity of the embedding functions \cite{janet,cartan} simplifies some results and imposes a maximum embedding dimension for all four-dimensional space-times to be $D= 10$. On the other hand, if one assumes that the deformed manifolds remain (at least) differentiable, the limit dimension for flat embeddings rises to $D= 14$ (such a number may be interesting for the study of super-algebras as in \cite{bars}), with a wide range of compatible signatures as shown by Greene \cite{Greene} extending Nash theorem's results for a $D$-dimensional bulk $n(n+ 3)/2$, as $n$ refers to the dimension of the embedded space. Also, $D= 14$ may induce a GUT based on a 45 parameter group like for example SO(10) or equivalent, proposing that a larger gauge symmetry may be possible to obtain.

As shown, the Nash-Greene theorem constitutes an important improvement over the traditional analytic embedding theorems \cite{janet,cartan}. Thus, our present structure fulfills such requirement that limits the number of the extra-dimen-\\sions up to 10. Such limitation avoids bulk instabilities and the appearance of ghosts like those of some brane-world models. For instance, one of the points of ref.\cite{luty} lies in the application of the longitudinal Goldstone mode acting as a brane bending mode. This was stated to keep fixed the induced metric on the brane. Unfortunately, this generates classical instabilities (negative energy solutions) of this type of DGP model. In the realm of brane-world program, it has been implemented mostly on particular models where the bulk has a fixed geometry and the space-time has a specific metric \emph{ansatz}. Such new geometric quantity has a paramount role on the dynamics of the embedding itself. Then, the extrinsic curvature is not replaced by any additional principle or algebraic relation (e.g., Israel-Lanczos condition). The ``bending-stretching'' (and eventually, the ``ripping'') of the geometry is a natural effect of the extrinsic curvature once the embedding is properly defined.

It is important to point out that the  confinement  of  the  gauge  fields to  four  dimensions is  not   really  an  \emph{ad-hoc} assumption. It  is  a  consequence  of the fact that  only  in  four  dimensions  the  three-form  resulting  from the  derivative of  the Yang-Mills  curvature   tensor  is  isomorphic  to the  one-form current. Consequently,  all  known  observable  sources  of gravitation  composing   the stress energy momentum tensor of the bulk $T_{AB}$ are   confined to  these   deformable    four-dimensional  space-times,    independently of  the  value  of the  extra  coordinate  $y$. Unless experimental evidences prove the contrary, the confinement hypothesis imposes that the gauge interactions are only restricted to the embedded space-time, even though a mathematically constructed higher dimensional extensions of Yang-Mills equations are always possible in the context of strings and branes. Moreover, the  confined   components of  $\mathcal{T}_{AB}$  can be proposed in the perturbed Gaussian frame $\{\mathcal{Z}^A_{,\mu},\eta^{B}\}$ as proportional to  the  energy-momentum tensor in such a way:
\begin{eqnarray}
&&T^{\ast}_{\mu\nu}=\;T^{\ast}_{AB}\mathcal{Z}^A_{,\mu}\mathcal{Z}^B_{,\nu}\;, \label{eq:confined1}\\
&&T^{\ast}_{\mu b}=\;T^{\ast}_{AB}\mathcal{Z}^A_{,\mu} \eta^B\;, \label{eq:confined2} \\
&&T^{\ast}=\;T^{\ast}_{AB}\eta^A\eta^B\;,\label{eq:confined3}
\end{eqnarray}
Unlike RS models and variants such as the confinement condition commonly takes the form $\alpha^{\ast}T^{\ast}_{\mu\nu}=T_{\mu\nu}\delta(y)$, where $T_{\mu\nu}$ vanishes in extra-coordinate with a boundary $y=0$, in Nash dynamical embedding, the confinement must prevail in all points of the embedded space-time. As a consequence, we have the conditions
\begin{eqnarray}
&&\alpha_\ast T^{\ast}_{\mu\nu}=8\pi G T_{\mu\nu}\;, \label{eq:confined4}\\
&&\alpha_\ast T^{\ast}_{\mu }=\;0\;\;, \label{eq:confined5} \\
&&\alpha_\ast T^{\ast}=\;0\;\;.\label{eq:confined6}
\end{eqnarray}

Using the embedding equations in Eqs.(\ref{eq:imersao 1}), (\ref{eq:imersao 2}) and (\ref{eq:imersao 3}), and by direct calculation of Gauss equation in Eq.(\ref{eq:G1}) contracted with the metric $g_{\mu\nu}$ , they lead to
\begin{eqnarray}\label{eq:ricciimersao}
R_{\mu\nu}=\left(g^{\rho\sigma}k_{\mu\rho }k_{\nu\sigma }- k_{\mu\nu }h \right) +^5\mathcal{R}_{AB}\mathcal{Z}^A_{,\mu}\mathcal{Z}^B_{,\nu} \\ \nonumber
\hspace{1.5cm}-^5\mathcal{R}_{ABCD}\eta^{A}\mathcal{Z}^C_{,\nu}\mathcal{Z}^B_{,\mu}\eta^{D}\;. \end{eqnarray}

A further contraction with the metric $g_{\mu\nu}$  leads to an explicit  relation of the Ricci scalar of the bulk $^5\mathcal{R}$ with the embedded four dimensional Ricci scalar $R$ as
\begin{equation}\label{eq:escalar de curv imersao}
R=\left(K^2-h^2\right)+ ^5\mathcal{R}-2\;^5\mathcal{R}_{AB}\eta^{A}\eta^{B}+^5\mathcal{R}_{ABCD}\eta^{A}\eta^{B}\eta^{C}\eta^{D}\;.
\end{equation}
The last two terms  vanishes when using the confinement to the sooner, and the latter turns a redundant surface term, and one obtains the action in a form given by Eq.(\ref{eq:action2}). With Eqs.(\ref{eq:ricciimersao}) and (\ref{eq:escalar de curv imersao}), one writes
\begin{equation}
\begin{aligned}
R_{\mu\nu}-\frac{1}{2}Rg_{\mu\nu}={}&\left(-g^{\rho\sigma}k_{\mu\rho }k_{\nu\sigma}+ k_{\mu\nu }h \right) +^5\mathcal{R}_{AB}\mathcal{Z}^A_{,\mu}\mathcal{Z}^B_{,\nu}\\
                                  {}& \hspace{0.1cm}-^5\mathcal{R}_{ABCD}\eta^{A}\mathcal{Z}^C_{,\nu}\mathcal{Z}^B_{,\mu}\eta^{D}+\frac{1}{2}\left(K^2-h^2\right)g_{\mu\nu}\\
                                  {}& \hspace{0.1cm} -\frac{1}{2}^5\mathcal{R}g_{\mu\nu}+^5\mathcal{R}_{AB}\eta^{A}\eta^{B}g_{\mu\nu}\\
                                  {}& \hspace{0.1cm}-\frac{1}{2} g_{\mu\nu}^5\mathcal{R}_{ABCD}\eta^{A}\eta^{B}\eta^{C}\eta^{D},\;\label{eq:gtensor1}
\end{aligned}
\end{equation}
Then, from Eq.(\ref{eq:gtensor1}), one defines the deformation tensor $Q_{\mu\nu}$
\begin{equation}\label{eq:q mu nu}
  Q_{\mu\nu}\equiv\;\left(g^{\rho\sigma}k_{\mu\rho }k_{\nu\sigma}- k_{\mu\nu }h
\right)-\frac{1}{2}\left(K^2-h^2\right)g_{\mu\nu}\;\;,
\end{equation}
which is straightforwardly conserved by direct derivation such as
\begin{equation}\label{eq:qmunuconserved}
Q^{\mu\nu}_{\;\;;\nu}=0\;.
\end{equation}
It is important to point out that such property of the deformation tensor is valid for an arbitrary number od dimensions. After neglecting redundant surface terms, one obtains
\begin{equation}
\begin{aligned}
R_{\mu\nu}-\frac{1}{2}Rg_{\mu\nu}+Q_{\mu\nu}={}&\left(^5\mathcal{R}_{AB}-\frac{1}{2}\;^5\mathcal{R}\mathcal{G}_{AB}\right)
\mathcal{Z}^A_{,\mu}\mathcal{Z}^B_{,\nu}\\
                                         {}& \hspace{0.3cm}^5\mathcal{R}_{AB}\eta^{A}\eta^{B}g_{\mu\nu}.\;\label{eq:gtensor1I}
\end{aligned}
\end{equation}

Using the confinement conditions in Eq.(\ref{eq:confined4}),  then one has
$$\left(^5\mathcal{R}_{AB}-\frac{1}{2}\;^5\mathcal{R}\mathcal{G}_{AB}\right)
\mathcal{Z}^A_{,\mu}\mathcal{Z}^B_{,\nu}=\;\alpha_\ast
T^\ast_{AB}\mathcal{Z}^A_{,\mu}\mathcal{Z}^B_{,\nu}=\;\alpha_\ast
T^\ast_{\mu\nu}\;\;.$$
Hence, after using Eq.(\ref{eq:confined4}), we obtain the first induced field equation
\begin{equation}\label{eq:gravitensor}
R_{\mu\nu}-\frac{1}{2}Rg_{\mu\nu}+\;Q_{\mu\nu}= 8\pi GT_{\mu\nu}\;.
\end{equation}
A similar process is made using the Codazzi equation in Eq.(\ref{eq:C1}) contracted with $g^{\nu\rho}$
$$g^{\nu\rho}\;^5\mathcal{R}_{ABCD}\mathcal{Z}^A_{,\mu}\eta^B\mathcal{Z}^C_{,\nu}\mathcal{Z}^D_{,\rho}
=\;g^{\nu\rho}k_{\mu\nu ;\rho}-g^{\nu\rho}k_{\mu\rho;\nu}\;,$$
and after eliminating redundant surface terms, it leads to
\begin{equation}\label{eq:relacao 37}
^5\mathcal{R}_{AB}\mathcal{Z}^A_{,\mu}\eta^{B}=\;^5\mathcal{R}_{ABCD}\mathcal{Z}^A_{,\mu}\eta^{B}\eta^{C}\eta^{D}+ g^{\nu\rho}k_{\mu\nu;\rho}-g^{\nu\rho}k_{\mu\rho;\nu}\;.
\end{equation}
On the other hand, taking Einstein equations for the bulk in Eq.(\ref{eq:EEbulk}) in the Gaussian frame $\{\mathcal{Z}^A_{,\mu},\eta^{B}\}$ and taking Eq.(\ref{eq:confined2}),
one gets
$$^5\mathcal{R}_{AB}\mathcal{Z}^A_{,\mu}\eta^{B}_{a}=\;\frac{1}{2}^5\mathcal{R}\mathcal{G}_{AB}\mathcal{Z}^A_{,\mu}\eta^{B}_{a}
=\;\frac{1}{2}\mathcal{R}g_{\mu a}=0\;,$$
and we obtain the second field equation
\begin{equation}\label{eq:gravivector}
   k_{\mu[\nu;\rho]} = 0\;.
\end{equation}
In references \cite{maia2,GDE,QBW,gde2}, it is shown the results for the embedding of a four-dimensional space-time in a D-dimen-sional space-time.  It is important to point out that Nash-Greene theorem is only valid for pseudo-Riemannian metric, but an interesting procedure should be in replacing the pseudo-Riemannian metric on which GR is based by a Finslerian metric. Physical models on such background have been explored in recent literature \cite{ikeda,konitopoulos} and show interesting consequences in amplifying  geometrical analysis and relations. As compared with our present model, we share a similar inner proposal that lies in expanding the concept of curvature. As a consequence, extra terms in the modified Friedmann equations leads to a novel approach to tackle the dark energy sector. In ref.\cite{konitopoulos}  was found that a quintessence or a phantom fluid behaviour is preferred, which is a similar result that our model proposes. Concerning embeddings, a somewhat similar development to Nash-Greene theorem was made by Burago and Ivanov \cite{burago} that showed  that any compact Finsler manifold can be isometrically embedded into a finite-dimensional normed space, which may be an interesting direction to follow in order to attain the physical consequences of such models.

\section{Background FLRW metric}
After establishing the induced field equations, we study the cosmological consequences in both background and perturbed universe. It is important to point out that the integrability conditions of the embedding (i.e., Gauss, Codazzi and Ricci equations), solved by Nash-Greene theorem by means of differentiable functions, reconstruct the five dimensional geometry and allow us to understand bulk properties from the perturbations of the four-dimensional space-time $V_4$ and vice-versa. The Riemann curvature of the bulk space acts as a reference for the Riemann curvature of the embedded space-time. Hence, the dynamics of the embedded four-dimensional space-time will lead the overall propagation of gravitational field to the bulk by means of the induced gravitational equations (from the bulk to the embedded space-time) leaving the Yang-Mills gauge fields confined to the inner space even under perturbations according to Nash-Greene theorem. Thus, we begin with the background cosmology and we rewrite the induced four dimensional equations for the background in an appropriate form. Rewriting Eqs.(\ref{eq:gravitensor}) and (\ref{eq:gravivector}) in a form
\begin{eqnarray}
  G^{(0)}_{\mu\nu}+ Q^{(0)}_{\mu\nu} &=&8\pi G T^{(0)}_{\mu\nu}\;, \label{eq:noperttensoreq}\\
  k^{(0)}_{\mu[\nu;\rho]} &=& 0 \label{eq:nopervecteq}\;,
\end{eqnarray}
where $ G^{(0)}_{\mu\nu}$ is the non-perturbed Einstein tensor, $T^{(0)}_{\mu\nu}$ denotes the non-perturbed energy-momentum tensor of the confined perfect fluid and $G$ is the gravitational Newtonian constant. For notation sake, we also rewrite Eq.(\ref{eq:q mu nu}) as the non-perturbed extrinsic term in a form $Q^{(0)}_{\mu\nu}$ in Eq.(\ref{eq:noperttensoreq}) and is given by
\begin{equation}\label{eq:qmunu}
  Q^{(0)}_{\mu\nu}=k^{(0)\rho}_{\mu}k^{(0)}_{\rho\nu}- k^{(0)}_{\mu\nu }h -\frac{1}{2}\left(K^2-h^2\right)g^{(0)}_{\mu\nu}\;,
\end{equation}
where we denote in this notation the mean curvature $h^{2\;(0)}=h^{(0)}\! \cdot \! h^{(0)}$ and $h= \;^0g^{\mu\nu}\;k^{(0)}_{\mu\nu}$. The term $K^{2\;(0)}=k^{\mu\nu\;(0)}k^{(0)}_{\mu\nu}$ is the Gaussian curvature. As shown in Eq.(\ref{eq:qmunuconserved}), the equation (\ref{eq:qmunu}) is readily conserved in the sense that
\begin{equation}\label{eq:qmunuconserv}
  Q^{(0)}_{\mu\nu;\mu}=0\;.
\end{equation}

As it happens, the basic familiar line element of FLRW is given by
\begin{equation}
  ds^2= -dt^2 + a^2\left(dr^2+r^2d\theta^2 + r^2\sin^2\theta d\phi^2 \right)\;,
\end{equation}
where the expansion factor is denoted by $a\equiv a(t)$. The coordinate $t$ denotes the physical time. In the Newtonian frame, the former equations turn out to be
\begin{equation}\label{eq:flrwnewtframe}
  ds^2= -dt^2 + a^2\left(dx^2+dy^2 + dz^2 \right)\;.
\end{equation}

\subsection{Non perturbed field equation in an embedded space-time}\label{noperturb}
By direct calculation of Eq.(\ref{eq:flrwnewtframe}) in Eq.(\ref{eq:noperttensoreq}), the components of $G^{(0)}_{\mu\nu}$ are given as usual \cite{kodasak, mukhanov}:
\begin{eqnarray}
&&G^{(0)}_{ij} = \frac{1}{a^2}\left(H^2+2\dot{H}\right)\delta_{ij}\;,\\
&&G^{(0)}_{4j} = 0\;,\\
&&G^{(0)}_{44} = \frac{3}{a^2} H^2\;,
 \end{eqnarray}
where the Hubble parameter is defined in the standard way by $H\equiv H(t)=\frac{\dot{a}}{a}$.

Since the extrinsic curvature is diagonal in FLRW space-time, one can find the components of extrinsic curvature using Eq.(\ref{eq:nopervecteq}) that can be split into spatial and time parts:
\begin{equation}
k^{(0)}_{ij,k}-\Gamma^{a}_{ik}k^{(0)}_{aj}= k^{(0)}_{ik,j}-\Gamma^{a}_{ij}k^{(0)}_{ak}\;.
\end{equation}
In the Newtonian frame, the spatial components are also symmetric and using the former relation one can obtain $k^{(0)}_{11}=k^{(0)}_{22}=k^{(0)}_{33}=b\equiv b(t)$, and straightforwardly
\begin{eqnarray}\label{eq:extcurvcomponents}
k^{(0)}_{ij}=\frac{b}{a^2}g_{ij},\;\;i,j=1,2,3, \;
k^{(0)}_{44}=\frac{-1}{\dot{a}}\frac{d}{dt}\frac{b}{a},
\end{eqnarray}
and the following objects can be determined as:
\begin{eqnarray}
\label{eq:BB}
 &&
 k^{(0)}_{44}=\frac{b}{a^{2}}\left(\frac{B}{H}-1\right)\;, \\
&&K^{2\;(0)}=\frac{b^2}{a^4}\left( \frac{B^2}{H^2}-2\frac BH+4\right),
 \;\, h^{(0)}=\frac{b}{a^2}\left(\frac BH+2\right),\label{eq:hk}\\
&&Q^{(0)}_{ij}= -\frac{1}{3}Q^{(0)}_{44}\left( 2\frac{B}{H}-1\right)\delta^{44}g^{(0)}_{ij},\;Q^{(0)}_{44} = -\frac{3b^{2}}{a^{4}},
  \label{eq:Qab}\\
&&Q^{(0)}= -(K^{2\;(0)} -h^{2\;(0)}) =\frac{6b^{2}}{a^{4}} \frac{B}{H}\;, \label{Q}
 \end{eqnarray}
where we define the function $B=B(t)\equiv \frac{\dot{b}}{b}$ in analogy with the Hubble parameter. At a time $t_0$, we define the dimensionless parameter $B|_{t=t_0}\equiv B_0$ and $b(t=t_0)\equiv b_0$. The parameter $B$ inherits $H$ dimensionality. The former results for background were originally obtained in \cite{GDE}.

\subsection{The Einstein-Gupta equations}
In  each    embedded  space-time  obtained by  the  smooth deformations,  the   metric  and the extrinsic  curvature  are   independent variables as shown by Nash-Greene theorem (Eq.(\ref{eq:nashdeformation})) in which does not provide \emph{per se} a dynamics for $k_{\mu\nu}$.  From the total of 20 unknowns  $g_{\mu\nu}$   and  $k_{\mu\nu}$, we count from Eq.(\ref{eq:EEbulk}) only   15  dynamical  equations. To solve that, the  remaining unknown equations come  from the  fact   that $k_{\mu\nu}$  is an  independent  symmetric  rank-2  tensor  that  corresponds to a  spin-2 field.  A theorem due  to   S. Gupta \cite{Gupta} shows  that such kind of tensors  necessarily  satisfy an  Einstein-like  system of  equations, having  the  Pauli-Fierz  equation  as its linear  approximation \cite{rham, Fronsdal}. Hence, to our purposes, we adapt Gupta's theorem to $k_{\mu\nu}$ to obtain its dynamics as originally shown in \cite{gde2}.

Suchlike to a projective space, we derive   Gupta's  equation   for  the    extrinsic  curvature  in the embedded space-time  endowed with  the metric  $^{(0)}g_{\mu\nu}$,  analogously to the standard derivation  of  Einstein's  equations. Then, we normalise the background extrinsic curvature by  noting that  $k^{(0)}_{\mu\nu}k^{\mu\nu\;(0)}  =K^2  \neq  4$, with the definition of  tensor
\be
f^{(0)}_{\mu\nu} = \frac{2}{K^{(0)}}k^{(0)}_{\mu\nu},
\label{eq:fmunu}
\ee
with its inverse by $ f^{\mu\rho\;(0)}f^{(0)}_{\rho\nu} =   \delta^\mu_\nu$.  Immediately, one obtains    $f^{\mu\nu\;(0)}=\frac{2}{K^{(0)}}k^{\mu\nu\;(0)}$. Denoting  by  $||$  the covariant derivative with respect to   a  connection  defined by $f^{(0)}_{\mu\nu}$,  while  keeping the usual semicolon notation for the covariant derivative with respect to $^{(0)}g_{\mu\nu}$,  the   analogous to the  ``Levi-Civita"  connection associated with $f^{(0)}_{\mu\nu}$  such   that ''  $f^{(0)}_{\mu\nu||\rho}=0$,  is:
\be
\Upsilon_{\mu\nu\sigma}=\;\frac{1}{2}\left(\partial_\mu\; f_{\sigma\nu}+ \partial_\nu\;f_{\sigma\mu} -\partial_\sigma\;f_{\mu\nu}\right) \;. \label{eq:upsilon}
\ee
Defining $\Upsilon_{\mu\nu}{}^{\lambda}= f^{\lambda\sigma\;(0)}\;\Upsilon^{(0)}_{\mu\nu\sigma}$, it allows to write the  ``Riemann tensor'' for  $f^{(0)}_{\mu\nu}$ that has  components
\be
{\cal  F}^{\;(0)}_{\nu\alpha\lambda\mu}= \;\partial_{\alpha}\Upsilon_{\mu\lambda\nu}- \;\partial_{\lambda}\Upsilon_{\mu\alpha\nu}+ \Upsilon_{\alpha\sigma\mu}\Upsilon_{\lambda\nu}^{\sigma} -\Upsilon_{\lambda\sigma\mu}\Upsilon_{\alpha\nu}^{\sigma}\;,
\label{eq:riemanntensorfmunu}
\ee
that leads to  the  corresponding  ``f-Ricci tensor'' and the ``f-Ricci scalar'' for  $f_{\mu\nu}$ written as,
$
{\cal  F}^{(0)}_{\mu\nu} =  f^{\alpha\lambda\;(0)}{\cal  F}^{(0)}_{\nu
\alpha\lambda\mu}$
and ${\cal  F}^{(0)}=f^{\mu\nu\;(0)}{\cal  F}^{\;(0)}_{\mu\nu}$, respectively. Finally,  Gupta's equations for $f^{(0)}_{\mu\nu}$  can be  obtained  from the  contracted  Bianchi  identity as
\begin{equation}
{\cal  F}^{(0)}_{\mu\nu}-\frac{1}{2}{\cal  F}^{(0)} f^{(0)}_{\mu\nu}=\;\alpha_*  \mathcal{T}_{\mu\nu}\;,\label{eq:gupta}
\end{equation}
where $\mathcal{T}_{\mu\nu}$ stands for the source of this field  such that   $\mathcal{T}^{\mu\nu}{}_{||\nu}=0$ and   $f_*$  is  a  coupling  constant. In  spite  of  the evident resemblances,    $k^{(0)}_{\mu\nu}$ is  not  a  metric tensor because it exists  only  after  the  Riemannian  geometry  has  been primely defined  with  the  metric  $^{(0)}g_{\mu\nu}$. Moreover, considering the simplest option for the ``f-curvature'', the related cosmological constant is zero and the only option for the  external  source  of  Eq.\rf{gupta}  is  the  void   characterized  by  $\mathcal{T}_{\mu\nu}=0$, once in this region, we only have pure gravitational interaction. With  such  interpretation,  Eq.\rf{gupta}  becomes   simply  a  Ricci-flat-like  equation
\be
{\cal  F}^{(0)}_{\mu\nu}=0 \;.  \label{eq:guptaflat}
\ee
Moreover, we use the spatially flat FLRW metric in Eq.(\ref{eq:flrwnewtframe}) and the definitions of Eqs.(\ref{eq:fmunu}) and (\ref{eq:upsilon}) to obtain
\begin{equation}
f^{(0)}_{ij}  =\frac{2}{K^{(0)}}\;^{(0)}g_{ij},\; i,j  = 1..3, \;   f^{(0)}_{44}  = -\frac{2}{K^{(0)}}\frac{1}{\dot{a}}\frac{d}{dt}{\left(\frac{b}{a}\right)}\;,
\label{eq:fij}
\end{equation}
from  which  we  derive  the  components of Eq.(\ref{eq:upsilon}) and
the  ``f-curvature'' ${\cal F}^{(0)}_{\mu\nu\rho\sigma}$. In this particular example, one obtains the components of Ricci-flat  equation in Eq.(\ref{eq:guptaflat}) given by
\begin{equation}
\begin{aligned}
\mathcal{F}^{(0)}_{11}={}&\frac{1}{4} \frac{-4b^2\xi\dot{K}^2 + 5b\xi\dot{K}\dot{b}K-\dot{b}^2\xi K^2 + 2b^2\xi K \ddot{K} }{\xi^2 K^2 b}\\
                                  {}& \hspace{0.5cm}+\frac{1}{4} \frac{ -2b\ddot{b}\xi K^2 - b^2\dot{K}\dot{\xi}K + bK^2\dot{b}\dot{\xi}}{\xi^2 K^2 b}=0\label{eq:11}
\end{aligned}
\end{equation}

\begin{equation}
\begin{aligned}
\mathcal{F}^{(0)}_{22}={}r^2 \frac{-4b^2\xi\dot{K}^2 + 5b\xi\dot{K}\dot{b}K-\dot{b}^2\xi K^2 + 2b^2\xi K \ddot{K}}{4\xi^2 K^2 b}\\
                                 {} + r^2 \frac{- 2b\ddot{b}\xi K^2 - b^2\dot{K}\dot{\xi}K + bK^2\dot{b}\dot{\xi}}{4\xi^2 K^2 b}=0
\end{aligned}
\end{equation}

\begin{equation}
\mathcal{F}^{(0)}_{33}=\sin^2(\theta)\mathcal{F}_{22}=0\;,
\end{equation}

\begin{equation}
\begin{aligned}
\mathcal{F}^{(0)}_{44}={}-3/4 \frac{\dot{b}^2\xi K^2 + 2b^2\xi K \ddot{K} -2b\ddot{b}\xi K^2 - b^2\dot{K}\dot{\xi}K }{\xi K^2  b^2}\\
                                  {}-3/4 \frac{b K^2\dot{b}\dot{\xi} -2b^2\xi \dot{K}^2 + b \xi K \dot{K}\dot{b}}{\xi K^2  b^2} =0, \label{eq:44}
\end{aligned}
\end{equation}
where for the sake of notation, we have denoted $\xi=k^{(0)}_{44}$ and retract the $^{(0)}$ in the squared root of Gaussian curvature $K$. By subtracting the relevant equations, i.e., Eq.(\ref{eq:11}) and Eq.(\ref{eq:44}), we obtain $b^2\dot{K}^2 + K^2\dot{b}^2 = 2bK\dot{b}\dot{K}$ or, equivalently,
\be
\left(\frac{\dot{K}}{K}\right)^2 - 2\frac{\dot{b}}{b} \frac{\dot{K}}{K} = -\left(\frac{\dot{b}}{b}\right)^2\;,
\ee
which  has  a   simple  solution   $K(t)= \sqrt{3}\eta_0 b(t)$, where the integration  constant was conveniently adjusted to $\sqrt{3}\eta_0$. Replacing the  expression of $K\equiv K(t)$ given by Eq.(\ref{eq:hk}), we obtain
\begin{equation}
\label{eq:EG1}
\frac{B}{H}=1\pm \sqrt{3}\sqrt{\eta_0^2 a^4 - 1}\;.
\end{equation}
In this form, Eq.(\ref{eq:EG1}) contributes to the Friedman equation as a dark radiation-like term. An amplification of this contribution can be done by simply adding an arbitrary dimensionless constant $c_0$ to Eq.(\ref{eq:EG1}). It results in a vertically shifted parent function of Eq.(\ref{eq:EG1}). Thus, we will have an enlargement of the range (image) of Eq.(\ref{eq:EG1}) but preserving its domain unchanged. From a physical perspective, it allows an additional contribution to Friedman equation leading to an accelerated expansion regime. As a result, one rewrites Eq.(\ref{eq:EG1}) in a generic form
\begin{equation}
\label{eq:EG2}
\frac{B}{H}=\beta_0\pm \sqrt{3}\sqrt{\eta_0^2 a^4 - 1}\;,
\end{equation}
where the summation $(1+c_0)$ is replaced by a dimensionless parameter $\beta_0$. Clearly, the parental function in Eq.(\ref{eq:EG2}) does not affect the evolution of $\left(\frac{B}{H}\right)$ in its time derivatives at any time $t$, at any order. In the limit $\beta_0\rightarrow 1$, one recovers Eq.(\ref{eq:EG1}). Moreover, the $\beta_0$ parameter must obey the constraint $\beta_0 \geq 1$ in order to be consistent with the Null energy condition (NEC) and should be constrained to data.

In addition, Eq.(\ref{eq:EG2}) can be readily integrated and one obtains the following $b(t)$ function as
\begin{equation}
b(t)  =  \left(\frac{b_0}{a_0^{\beta_0}}\exp{(\pm \gamma(a_0))}\right) a^{\beta_0} \cos\left(\frac{\sqrt{3}}{2}\sqrt{1-\eta_0^2a^4}\right) e^{\pm \gamma (a)}  \label{eq:b1}
\end{equation}
where $\gamma(a)$ is a complex function given  by
\begin{equation}
\!\!\gamma(a) = 
\! \!\arctan\left(i\sqrt{1-\eta_0^2 a^4}\right)\;,\label{eq:gamma}
\end{equation}
and so is $\gamma(a_0)=\gamma(a)|_{a=a_0}$. Calculating the modulus of the complex function $b(t)$, one obtains
\begin{equation}
|b(t)| =  \left(\frac{b_0}{a_0^{\beta_0}}\right) a^{\beta_0} \cos\left(\frac{\sqrt{3}}{2}\sqrt{1-\eta_0^2a^4}\right)\;.
\label{eq:b2}
\end{equation}
The square root term in Eq.(\ref{eq:b2}) also poses a bound over $\eta_{0}.$ The $\eta_0$ parameter is bounded from below at $a=a_0$ in such a way $\eta_{0}\leq 1$. Hereon, for the sake of notation, we simply write the modulus $|b(t)|$ as $|b(t)| \equiv b(t)$.

It is worth noting that the cosine function acts as a damping function on the evolution of Eq.(\ref{eq:b2}). We verified that a better growth behaviour is obtained when $\eta_0 \rightarrow 0$ for any $a(t)$ and $b(t)$ increases monotonically as
\begin{equation}
b(t)  =  \left(\frac{b_0}{a_0^{\beta_0}}\right) a^{\beta_0}\;,\label{eq:b3}
\end{equation}
which can be useful to mimic smooth dark energy perturbations. For the present time, $a_0$ can be set as $a_0=1$.

\subsection{Hydrodynamical equations}
The stress energy tensor in a non-perturbed co-moving fluid is given by
$$T^{(0)}_{\mu\nu}=\left(\rho^{(0)}+ p^{(0)}\right) u_{\mu}u_{\nu}+ p^{(0)}g^{(0)}_{\mu\nu}\;;\;u_{\mu}=\delta^4_{\mu}\;.$$
The conservation of $T^{(0)}_{\mu\nu;\mu}=0$ leads to
\begin{equation}\label{eq:nonpertdensity}
  \rho^{(0)} + 3H\left(\rho^{(0)} + p^{(0)}\right)=0\;.
\end{equation}
From the induced field equations in Eqs.(\ref{eq:noperttensoreq}) and  (\ref{eq:nopervecteq}), the resulting Friedmann equation turns
\begin{equation}\label{eq:nonpertfriedmann}
  H^2=\frac{8}{3}\pi G \rho^{(0)}+ \frac{b^2}{a^4}\;,
\end{equation}
where $\rho^{(0)}$ is the present value of the non-perturbed matter density (hereon $\rho^{(0)}\equiv \rho^{(0)}_{m}(t)$) and $b(t)$ is given by Eq.(\ref{eq:b3}). Thus, we write the matter density in terms of redshift as
\begin{equation}\label{eq:matdensitynonpert}
\rho^{(0)}_{m}(t)=\rho^{(0)}_{m(0)}a^{-3}=\rho^{(0)}_{m(0)}(1+z)^3\;.
\end{equation}
and we rewrite Eq.(\ref{eq:nonpertfriedmann}) in terms of redshift as
\begin{equation}\label{eq:nonpertfriedtotal2}
  H^2=\frac{8}{3}\pi G \rho^{(0)}_{m(0)}(1+z)^3 + b^2_0(1+z)^{4-2\beta_0}\;.
\end{equation}
Using the definition of the cosmological parameter $\Omega_{i}=\frac{8\pi G}{3H_0^2}\rho^{(0)}_{i(0)}$, we finally have
\begin{equation}\label{eq:nonpertfriedtotal3}
  \left(\frac{H}{H_0}\right)^2=\Omega_{m(0)}(1+z)^3 + (1- \Omega_{m(0)})(1+z)^{4-2\beta_0}\;,
\end{equation}
where $\Omega_{m(0)}$ is the current cosmological parameter for matter content and for a flat universe $\Omega_{ext(0)}= 1- \Omega_{m(0)}$. The extrinsic cosmological parameter $\Omega_{ext(0)}$ was written using a fluid analogy, in such a way we define
\begin{equation}\label{eq:extomega}
\Omega_{ext(0)}=\frac{8\pi G}{3H_0^2}\rho^{(0)}_{ext(0)}\equiv \frac{b_0^2}{a_0^{\beta_0}}\;.
\end{equation}
As standard practice, $H_0$ is the current value of Hubble constant in units of km.s$^{-1}$ Mpc$^{-1}$. Likewise, in conformal time $\eta$ such that $dt=a(\eta)d\eta$ and $\mathcal{H}\equiv aH$, we can write the Friedmann equation in this frame as
\begin{equation}\label{eq:nonpertfriedconformal}
  \mathcal{H}^2=\frac{k_0}{3}a^2 \left(\rho^{(0)}_m(t) + \frac{b^2_0}{k_0} a^{2\beta_0-4}\right)\;,
\end{equation}
where $k_0\equiv \frac{8}{3}\pi G$ and the conformal Hubble parameter is $\mathcal{H}=\frac{a'}{a}$. The prime symbol represents the conformal time derivative. Hence, the conformal time derivative of Hubble parameter is given by
\begin{equation}\label{hubbleconformaltimederivat}
\mathcal{H}'\equiv \frac{d\mathcal{H}}{d\eta}=-\frac{k_0}{6}a^2\left( \rho^{(0)}_{m} + 3p^{(0)} + (\beta_0-4)\frac{b^2_0}{k_0} a^{2\beta_0-4} \right) \;,
\end{equation}
and completes the set of equations for a non-perturbed fluid in a conformal Newtonian gauge.

\section{Transformations and gauge variables}
Using the standard line element of FLRW metric in Euclidean coordinates in Eq.(\ref{eq:flrwnewtframe}), one finds
\begin{equation}\label{eq:friedmaneuclidean}
ds^2=a^2\left(-d\eta^2 + \delta_{ij}dx^i dx^j\right)\;,
\end{equation}
where $a=a(\eta)$ is the expansion parameter in conformal time. We start with the standard process as known in GR \cite{kodasak, mukhanov,muk}. The novelty of this approach is the inclusion of the extrinsic curvature in the theoretical framework. Thus, let be the coordinate transformation $x^{\alpha}\rightarrow \tilde{x}^{\alpha}=x^{\alpha}+\xi^{\alpha}$ such as $\xi^{\alpha}\ll 1$, then we have for a second order tensor
$$\tilde{g}_{\alpha\beta}(\tilde{x}^{\rho})=   \frac{\partial x^{\gamma}}{\partial\tilde{x}^{\alpha}} \frac{\partial x^{\delta}}{\partial \tilde{x}^{\beta}}\;g_{\gamma\delta}(\tilde{x}^{\rho})\;.$$
Hence, we can write the perturbed metric tensor in the new coordinates $\delta \tilde{g}_{\alpha\beta}$ as
\begin{equation}\label{ctransform}
\delta \tilde{g}_{\alpha\beta}= \delta g_{\alpha\beta}- g^{(0)}_{\alpha\beta,\gamma}\xi^{\gamma}-g^{(0)}_{\alpha\delta}\xi^{\delta}_{,\beta}-g^{(0)}_{\beta\delta}\xi^{\delta}_{,\alpha}\;,
\end{equation}
where the infinitesimally vector function $\xi^{\alpha}=\xi^{(4)}+\xi^{i}$ can be split into two parts
$$\xi^{i}=\xi^{i_{\bot}}+ \zeta^{,i}\;,$$
in which $\xi^{i_{\bot}}$ is the orthogonal part decomposition and $\zeta$ is a scalar function. The prime symbol denotes the derivative with respect to conformal time $\eta$. As a result, we can obtain
\begin{eqnarray}
\delta \tilde{g}_{ij}&=& \delta g_{ij} + a^2\left[2\frac{a'}{a}\delta_{ij}\xi^{(4)} + 2\zeta_{,ij} + \xi^{\bot}_{i,j} + \xi^{\bot}_{j,i}\right]\;,\\ \label{ij}
\delta \tilde{g}_{4i}&=& \delta g_{4i} + a^2 \left( \xi'_{\bot i} + \left[\zeta'-\xi^{(4)}\right]_{,i} \right)\;, \\ \label{zeroi}
\delta \tilde{g}_{44}&=& \delta g_{44} - 2a(a\xi^{(4)})'\;. \label{zerocoord}
\end{eqnarray}

Using Eq.(\ref{ctransform}), we obtain a similar transformation for $k_{\mu\nu}$ as
\begin{equation}\label{ktransform}
\delta \tilde{k}_{\alpha\beta}= \delta k_{\alpha\beta}- k^{(0)}_{\alpha\beta,\gamma}\xi^{\gamma}-k^{(0)}_{\alpha\delta}\xi^{\delta}_{,\beta}-k^{(0)}_{\beta\delta}\xi^{\delta}_{,\alpha}\;.
\end{equation}
And taking into account the Nash-Greene theorem
\begin{equation}\label{nashtheor}
k^{(0)}_{\mu\nu}=-\frac{1}{2} g^{\bullet\;(0)}_{\mu\nu}\;,
\end{equation}
where we denote $g^{\bullet\;(0)}_{\mu\nu}= \frac{\partial g^{(0)}_{\mu\nu}}{\partial y}$, and $y$ is the coordinate of direction of perturbations from the background to the extra-dimensions (in this case, just one extra-dimension). Thus, we can rewrite Eq.(\ref{ktransform}) as
\begin{equation}\label{k2transform}
\delta \tilde{k}_{\alpha\beta}= \delta k_{\alpha\beta}+ \frac{1}{2} g^{\bullet\;(0)}_{\alpha\beta,\gamma}\xi^{\gamma}+\frac{1}{2} g^{\bullet\;(0)}_{\alpha\delta}\xi^{\delta}_{,\beta}+\frac{1}{2} g^{\bullet\;(0)}_{\beta\delta}\xi^{\delta}_{,\alpha}\;,
\end{equation}
and we get straightforwardly
\begin{eqnarray*}
\delta \tilde{k}_{ij}&=& \delta k_{ij} -(a^2)^{\bullet}\left[\frac{1}{2}\frac{\left((a^2)^{\bullet}\right)_{,4}}{(a^2)^{\bullet}}\delta_{ij}\xi^{(4)} + 2\zeta_{,ij} + \xi^{\bot}_{i,j} + \xi^{\bot}_{j,i}\right]\;,\\
\delta \tilde{k}_{4i}&=& \delta k_{4i} + \frac{1}{2}(a^2)^{\bullet} \left( \xi'_{\bot i} + \left[\zeta'-\xi^{(4)}\right]_{,i} \right)\;,
 \\
\delta \tilde{k}_{44}&=& \delta k_{44} + \left((a^{2})^{\bullet}\xi^{(4)}\right)_{,4} - \frac{1}{2}\left((a^2)^{\bullet}\right)_{,4}\xi^{(4)}\;.
\end{eqnarray*}
Taking the previous expressions and to avoid the implications of the ambiguity of two ``times'' coordinates, likewise the Rosen bi-metric theory \cite{nrosen,will} that led to erroneous results such as a dipole gravitational waves, we adopt $\mathbf{y}$ as a set of space-like coordinates. Then, one obtains
\begin{eqnarray}
\delta \tilde{k}_{ij}&=& \delta k_{ij} \;, \\ \label{ijk}
\delta \tilde{k}_{4i}&=& \delta k_{4i} \;,  \\\label{zeroik}
\delta \tilde{k}_{44}&=& \delta k_{44} \;.  \label{zerocoordk}
\end{eqnarray}

For scalar perturbations the metric takes the form
\begin{equation}
\begin{aligned}
ds^2={}&a^2 [-(1+ 2\phi) d\eta^2 + 2B_{,i}dx^i d\eta\\
               &\hspace{1.5cm} + ((1-2\psi)\delta_{ij}- 2E_{,ij})dx^i dx^j] \;,\label{eq:scalarpertmetric}
\end{aligned}
\end{equation}
where $\phi=\phi(\vec{x},\eta)$, $\psi=\psi(\vec{x},\eta)$, $B=B(\vec{x},\eta)$ and $E=E(\vec{x},\eta)$ are scalar functions.

For the tensors $G_{\mu\nu}$, $T_{\mu\nu}$ and $Q_{\mu\nu}$, one can use the same set of transformations. In this sense, for small perturbations, we can write the Einstein tensor in a coordinate system $\tilde{x}$ as
$$\tilde{G}_{\mu\nu} =G^{(0)}_{\mu\nu} + \delta \tilde{G}_{\mu\nu}\;,$$
where $\delta \tilde{G}_{\mu\nu}$ denotes linear perturbations in the new coordinate system
\begin{equation}\label{Gtransform}
\delta \tilde{G}_{\alpha\beta}= \delta G_{\alpha\beta}- G^{(0)}_{\alpha\beta,\gamma}\xi^{\gamma}-G^{(0)}_{\alpha\delta}\xi^{\delta}_{,\beta}-G^{(0)}_{\beta\delta}\xi^{\delta}_{,\alpha}\;.
\end{equation}
Immediately, we have a similar expression for $T_{\mu\nu}$
$$\tilde{T}_{\mu\nu} =T^{(0)}_{\mu\nu} + \delta \tilde{T}_{\mu\nu}\;,$$
that leads to
\begin{equation}\label{Ttransform}
\delta \tilde{T}_{\alpha\beta}= \delta T_{\alpha\beta}- T^{(0)}_{\alpha\beta,\gamma}\xi^{\gamma}-T^{(0)}_{\alpha\delta}\xi^{\delta}_{,\beta}-T^{(0)}_{\beta\delta}\xi^{\delta}_{,\alpha}\;,
\end{equation}
and also, for the deformation tensor
$$\tilde{Q}_{\mu\nu} =Q^{(0)}_{\mu\nu} + \delta \tilde{Q}_{\mu\nu}\;,$$
that leads to
\begin{equation}\label{Qtransform}
\delta \tilde{Q}_{\alpha\beta}= \delta Q_{\alpha\beta}- Q^{(0)}_{\alpha\beta,\gamma}\xi^{\gamma}-Q^{(0)}_{\alpha\delta}\xi^{\delta}_{,\beta}-Q^{(0)}_{\beta\delta}\xi^{\delta}_{,\alpha}\;.
\end{equation}
And using Eq.(\ref{eq:scalarpertmetric}), we obtain for $\delta \tilde{G}_{\mu\nu}$:
\begin{eqnarray}
  \delta \tilde{G}_{j}^{\;\;i} &=& \delta G_{j}^{\;\;i} - (^{(0)}G_{j}^{\;\;i})' (B-E')\;,\\
  \delta \tilde{G}_{i}^{\;\;4} &=& \delta G_{i}^{\;\;4} - (^{(0)}G_{4}^{\;\;4}- \frac{1}{3}\;^{(0)}G_{k}^{\;\;k}) (B-E')_{,i}\;,\\
  \delta \tilde{G}_{4}^{\;\;4} &=& \delta G_{4}^{\;\;4} - (^{(0)}G_{4}^{\;\;4})' (B-E')\;.
\end{eqnarray}
For the perturbed stress energy tensor $\delta \tilde{T}_{\mu\nu}$, one obtains the set of equations
\begin{eqnarray}
  \delta \tilde{T}_{j}^{\;\;i} &=& \delta T_{j}^{\;\;i} - (^{(0)}T_{j}^{\;\;i})' (B-E')\;\\
  \delta \tilde{T}_{i}^{\;\;4} &=& \delta T_{i}^{\;\;4} - (^{(0)}T_{4}^{\;\;4}- \frac{1}{3}\;^{(0)}T_{k}^{\;\;k}) (B-E')_{,i}\;,\\
  \delta \tilde{T}_{4}^{\;\;4} &=& \delta T_{4}^{\;\;4} - (^{(0)}T_{4}^{\;\;4})' (B-E')\;.
\end{eqnarray}
Likewise, for the perturbed induced extrinsic part $\delta \tilde{Q}_{\mu\nu}$ we have
\begin{eqnarray}
\delta \tilde{Q}_{j}^{\;\;i} &=& \delta Q_{j}^{\;\;i} - (^{(0)}Q_{j}^{\;\;i})' (B-E')\;\\
\delta \tilde{Q}_{i}^{\;\;4} &=& \delta Q_{i}^{\;\;4} - (^{(0)}Q_{4}^{\;\;4}- \frac{1}{3}\;^{(0)}Q_{k}^{\;\;k}) (B-E')_{,i}\;,\\
\delta \tilde{Q}_{4}^{\;\;4} &=& \delta Q_{4}^{\;\;4} - (^{(0)}Q_{4}^{\;\;4})' (B-E')\;.
\end{eqnarray}

\section{Scalar perturbations in Newtonian gauge}

\subsection{Perturbed gravitational equations}
In longitudinal conformal Newtonian gauge, the main condition resides in the vanishing functions of $B=B(\vec{x},\eta)$ and $E=E(\vec{x},\eta)$, as well as the quantities $\xi^{(4)}, \xi', \zeta$. Hence, the metric in Eq.(\ref{eq:scalarpertmetric}) turns to be
\begin{equation}\label{eq:scalarpertmetric2}
ds^2 = a^2 [(1+ 2\Phi) d\eta^2 - (1-2\Psi)\delta_{ij} dx^i dx^j] \;,
\end{equation}
where $\Phi=\Phi(\vec{x},\eta)$ and $\Psi=\Psi(\vec{x},\eta)$ denote the Newtonian potential and the Newtonian curvature, respectively. In addition, we obtain a simplification of the previous transformations of the curvature-related quantities and the set of following outcomes:
\begin{eqnarray}
&&\delta \tilde{g}_{44} =\delta g_{44}\;;\;\delta \tilde{g}_{4i}= \delta g_{4i}=0 \;;\;\delta \tilde{g}_{ij}= \delta g_{ij} \;, \label{eq:coordscalar} \\
&&  \delta \tilde{G}_{4}^{\;\;4} = \delta G_{4}^{\;\;4}\;;\;\delta \tilde{G}_{i}^{\;\;4} = \delta G_{i}^{\;\;4}\;;\;\delta \tilde{G}_{j}^{\;\;i} = \delta G_{j}^{\;\;i} \;, \\\nonumber
&&  \delta \tilde{T}_{4}^{\;\;4} = \delta T_{4}^{\;\;4}\;;\;\delta \tilde{T}_{i}^{\;\;4} = \delta T_{i}^{\;\;4}\;;\;\delta \tilde{T}_{j}^{\;\;i} = \delta T_{j}^{\;\;i} \;, \\ \nonumber
&&  \delta \tilde{Q}_{4}^{\;\;4} = \delta Q_{4}^{\;\;4}\;;\;\delta \tilde{Q}_{i}^{\;\;4} = \delta Q_{i}^{\;\;4}\;;\;\delta \tilde{Q}_{j}^{\;\;i} = \delta Q_{j}^{\;\;i} \;. \nonumber
\end{eqnarray}
Taking into account all the former results of Eqs.(\ref{eq:coordscalar}), we can write the perturbed induced field equations simply as
\begin{eqnarray}
&&\delta G^{\mu}_{\nu}=\;8\pi G\delta T^{\mu}_{\nu}-\delta Q^{\mu}_{\nu}\;, \label{eq:perturbgravitensor} \\
&&\delta k_{\mu\nu;\rho}=\; \delta k_{\mu\rho;\nu}\;.\label{eq:perturbcodazzi}
\end{eqnarray}
Using the Nash-Greene theorem, we notice that Codazzi equations in Eq.(\ref{eq:perturbcodazzi}) do not propagate perturbations in which are confined to the background. In other words, Codazzi equations maintain their background form and so are the perturbations of Gupta equations in which $\delta {\cal  F}_{\mu\nu}=0$.

Applying Eq.(\ref{perturbkmunu}) to Eq.(\ref{eq:perturbcodazzi}), we obtain the background equation as in Eq.(\ref{eq:nopervecteq}). In this sense, we have to look for the effects of the Nash-Greene fluctuations on the perturbed gravi-tensor equation in Eq.(\ref{eq:perturbgravitensor}) and verify if the propagations of cosmological perturbations may occur. Thus, we can write the components of Eq.(\ref{eq:perturbgravitensor}) as
\begin{eqnarray*}
\delta G^{i}_{j} &=& 8\pi G\delta T^{i}_{j} - \delta Q^{i}_{j} \;,\\
\delta G^{4}_{i} &=& 8\pi G\delta T^{4}_{i} - \delta Q^{4}_{i} \;, \\
\delta G^{4}_{4} &=& 8\pi G\delta T^{4}_{4} - \delta Q^{4}_{4} \;,
\end{eqnarray*}
and using the conformal metric in Eq.(\ref{eq:scalarpertmetric2}), we have the components in the conformal Newtonian frame,
\begin{eqnarray}\label{tensorcompij}
&&\mathcal{D}\delta_{ij}=\frac{1}{2} (\Psi-\Phi)_{,ij}+ \frac{1}{2} a^2 \delta Q^{i}_{j} - 4\pi G a^2 \delta T^{i}_{j}\;,\\
&&\left[\Psi'+ \mathcal{H}\Phi\right]_{,i}= 4\pi G a^2\delta T^{4}_{i} - \frac{1}{2} a^2\delta Q^{4}_{i}\;,\label{tensorcomp0i}\\
&&\nabla^2\Psi - 3 \mathcal{H} \left(\Psi'+ \Phi \mathcal{H} \right)= 4\pi G a^2 \delta T^{4}_{4} - \frac{1}{2}a^2 \delta Q^{4}_{4} \;.\label{tensorcompo00}
\end{eqnarray}
where $\mathcal{D}=\Psi'' + \mathcal{H}(2\Psi+\Phi)' + (\mathcal{H}^2+2\mathcal{H}')\Phi+ \frac{1}{2} \nabla^2(\Psi-\Phi)$. Actually, the former perturbed Einstein equations in con-formal-Newtonian gauge coincide with those written in an arbitrary coordinate system \cite{muk}, which naturally relates this gauge to a gauge-invariant framework. Thus, the perturbations $\delta k_{\mu\nu}$ can be determined by the metric perturbations as shown in Eq.(\ref{perturbkmunu}) in an adopted gauge. Moreover, the perturbation of the deformation tensor $Q_{\mu\nu}$ can be made from its background form in Eq.(\ref{eq:qmunu}) and the resulting $k_{\mu\nu}$ perturbations from the Nash fluctuations of Eq.(\ref{perturbkmunu}) such as
\begin{equation}\label{Qmunu}
\delta Q_{\mu\nu}= -\frac{3}{2}(K^2-h^2)\delta g_{\mu\nu} \;.
\end{equation}
The quantity $\delta Q_{\mu\nu}$ is also independently conserved in a sense that $\delta Q_{\mu\nu;\nu}=0$. Moreover, using the background relations of Eqs.(\ref{eq:BB}), (\ref{eq:hk}), (\ref{eq:Qab}), (\ref{Q}), we can determine the components of $\delta Q_{\mu\nu}$ as
\begin{eqnarray}
\delta Q^{i}_{j}&=& 18 \beta_0 b_0^2 a^{2\beta_0-2}\Psi \delta^i_j\;,\\
\delta Q^{i}_{4}&=& 0\;, \\
\delta Q^{4}_{4}&=& 18 \beta_0 b_0^2 a^{2\beta_0-2}\Phi \delta^4_4\;.
\end{eqnarray}
Thus, we get the basic gauge invariant field equations modified by the extrinsic curvature in the conformal Newtonian gauge as
\begin{eqnarray}
&&\mathcal{D}\delta_{ij}=9\gamma_0 a^{2\beta_0}\Psi \delta_{ij}+ \frac{1}{2} (\Psi-\Phi)_{,ij} - 4\pi G a^2 \delta T^{i}_{j}\;, \label{tensorcompij2} \\
&&\left[\Psi'+ \mathcal{H}\Phi\right]_{,i}= 4\pi G a^2\delta T^{4}_{i}\;, \label{tensorcomp0i2}\\
&&\nabla^2\Psi - 3 \mathcal{H} \left(\Psi'+ \Phi \mathcal{H} \right)= 4\pi G a^2 \delta T^{4}_{4} - 9\gamma_0 a^{2\beta_0}\Phi \;,\label{tensorcompo002}
\end{eqnarray}
where we denote $\gamma_0=\beta_0b_0^2$. Using Eq.(\ref{eq:extomega}), $\gamma_0$ parameter is simply written in terms of $\beta_0$ as
\begin{equation}\label{eq:gammaeq}
  \gamma_0=\beta_0\;\Omega_{ext(0)}\;,
\end{equation}
where $\Omega_{ext(0)}$ is the extrinsic (current) cosmological parameter.

\subsection{Hydrodynamical gravitational perturbed equations}
For a perturbed fluid with pressure $p$ and density $\rho$, one can write the perturbed components of the related stress-tensor
\begin{eqnarray}
&&  \delta \tilde{T^4_4} = \delta \rho\;, \\
&&  \delta \tilde{T^4_i} = \frac{1}{a}(\rho_0+p_0)\delta u_{\parallel i}\;, \\
&&  \delta \tilde{T^i_j} = -\delta p\;\delta^i_j\;,
\end{eqnarray}
where $\delta u_{\parallel i}$ denotes the tangent velocity potential and $\rho_0$ and $p_0$ denote the non-perturbed components of density and pressure, respectively. Hence, we can rewrite Eqs.(\ref{tensorcompij2}), (\ref{tensorcomp0i2}) and (\ref{tensorcompo002}) as
\begin{eqnarray}
&&\nabla^2\Psi - 3 \mathcal{H} \left(\Psi'+ \Phi \mathcal{H} \right)= 4\pi G a^2 \delta \rho - 9\gamma_0 a^{2\beta_0}\Phi\;, \label{tensorcompo00hidro} \\
&&\left[\Psi'+ \mathcal{H}\Phi\right]_{,i}= 4\pi G a(\rho_0+p_0) \delta u_{\parallel i}\;,\label{tensorcomp0ihidro}\\
&&\mathcal{D}\delta_{ij}-\frac{1}{2} (\Psi-\Phi)_{,ij}  = \left[4\pi G a^2 \delta p\;+9\gamma_0 a^{2\beta_0}\Psi\right] \delta_{ij}.\label{tensorcompijhidro}
\end{eqnarray}
These set of equations can be better understood in the Fourier \emph{k}-space wave modes.  In order to maintain the correct physical dimensions in the \emph{k}-space modes of objects from the extrinsic geometry, we need to project the perturbed $\delta Q_{\mu\nu}$ onto the intrinsic \emph{k}-wave vector field. Therefore, we make possible a further consistent cosmic fluid correspondence of extrinsic curvature, which is a fully geometric component. Thus, considering a vector field $\vec{k}= k^i \frac{\partial}{\partial w^i}$  with local coordinates $\{ w^1, w^2, ..., w^n\}$ and coordinate momenta $k^i= \frac{d w^i}{dt}$. Hence, the perturbed $\delta Q_{\mu\nu}$ can be dragged into the $\vec{k}$ vector flow by Lie derivative $\pounds_{\vec{k}}$. Then, one defines the Fourier transform
\begin{equation}\label{eq:fourierqmunu}
  \delta Q_{\mu\nu}= \int \pounds_{\vec{k}} \delta Q_{\mu\nu}^{(\vec{k})} (k,w,a) e^{i\vec{k}\bullet \vec{x}} dk\;,
\end{equation}
where $\delta Q_{\mu\nu}^{(\vec{k})} (k,w,a)$  is the defined in terms of variables $(k,w,a)$ as
\begin{equation}\label{eq:fourierqmunu2}
   \delta Q_{\mu\nu}^{(\vec{k})} (k,w,a)= \delta_{ij} k^i w^j f(a) \delta g_{\mu\nu}\;.
\end{equation}
where $f(a)$ is an \emph{ab-initio} arbitrary function of the expansion factor $a$.

In FLRW space-time,  $\vec{k}$ vector flow is conserved in a sense $k^{\rho}_{;\mu}= k^{\rho}_{;\nu}=0$, and one obtains the drag of $\delta Q_{\mu\nu}$ in k-space as
\begin{equation}\label{eq:fourierqmunu3}
  \delta Q_{\mu\nu}= \int  k^2 f(a) \delta g_{\mu\nu} e^{i\vec{k}\bullet \vec{x}} dk\;,
\end{equation}
where, in our present application, $f(a)= 9\gamma_0 a^{2\beta_0} $ .

Taking the Fourier transform of each main quantity (with subscript ``\emph{k}''), we obtain a new set of equations:
\begin{eqnarray}
&&k^2\Psi_k + 3 \mathcal{H} \left(\Psi^{'}_k + \Phi_k \mathcal{H} \right)= -4\pi G a^2 \delta \rho_k + \chi(a)\Phi_k , \label{tensorcompo00kspace}\\
&&\Psi^{'}_{k}+ \mathcal{H}\Phi_k= -4\pi G a^2(\rho_0+p_0) \theta\;,\label{tensorcomp0ikspace}
\end{eqnarray}
where $\theta= ik^j\delta u_{\parallel j}$ denotes the divergence of fluid velocity in \emph{k}-space and  $\chi(a)=9\gamma_0 k^2 a^{2\beta_0}$.  Finally, the third equation is given by
\begin{eqnarray}\label{tensorcompijkspace}
&&\mathcal{D}_k - \frac{1}{2}\hat{k}^i\cdot \hat{k}_i (\Psi_k-\Phi_k)= 4\pi G a^2 \delta p + \chi(a)\Psi_k\;,
\end{eqnarray}
where $\mathcal{D}_k=\Psi^{''}_{k} + \mathcal{H}(2\Psi_k+\Phi_k)' + (\mathcal{H}^2+2\mathcal{H}')\Phi_k+ \frac{1}{2} k^2(\Psi_k-\Phi_k)$.

\section{Matter density evolution under subhorizon regime}
In order to obtain a bare response of an influence of the extrinsic terms, we do not consider anisotropic stresses and pressure for Eq.(\ref{tensorcompo00kspace}), Eq.(\ref{tensorcomp0ikspace}) and Eq.(\ref{tensorcompijkspace}), we obtain the following equation
\begin{equation}\label{tensorcompo00kspace2}
k^2\Phi_k + 3 \mathcal{H} \left(\Phi^{'}_k + \Phi_k \mathcal{H} \right)= -4\pi G a^2 \delta \rho_k + \chi(a)\Phi_k \;,
\end{equation}
where the closure condition $\Psi=\Phi$ applies which is a result of the space-space traceless component from Eq.(\ref{tensorcompijhidro}).

It is important to notice that when $\gamma_0\rightarrow 0$ (i.e., when extrinsic curvature component vanishes $b_0\rightarrow 0$) in Eq.(\ref{tensorcompo00kspace2}), the standard GR equations are obtained and we recover the subhorizon approximation with $k^2\gg\mathcal{H}^2$ or $k^2\gg a^2H^2$ which means $\Phi''_k, \mathcal{H}\Phi'_k\sim 0$. To determine the gravitational potential $\Phi$, we also need to work with the continuity and Euler equations from calculating the components $\delta T^{\mu}_{\mu;4}=0$ and $\delta T^{\mu}_{\mu;i}=0$ to obtain, respectively
\begin{eqnarray}
&&  \delta \rho' +(p_0+\rho_0)\xi(\Phi)+ 3\mathcal{H}(\delta p+ \delta \rho)=0\;,\label{continuity} \\
&&  \frac{d}{d\eta}\left[(p_0+\rho_0)u_i\right]+ (p_0+\rho_0)\zeta(\Phi)+ \delta p=0\;, \label{euler}
\end{eqnarray}
where we denote $\xi(\Phi)=\nabla^2 u_i - 3 \Phi'$ and $\zeta(\Phi)=4\mathcal{H} u_i+\Phi$. Moreover, taking Eq.(\ref{continuity}) under a Fourier transform, we obtain the following equation in the \emph{k}-space as
$$\delta \rho'_k - k^2 \rho_0 u_k - 3\rho_0\Phi' + 3\mathcal{H}\delta \rho_k=0\;,$$
which in subhorizon approximation gives
\begin{equation}\label{denssubhoriz}
  \delta \rho'_k - k^2 \rho_0 u_k \simeq 0\;.
\end{equation}
For the pressureless form of Eq.(\ref{euler}), we have
$$\rho'_k u_k + \rho_k u'_k + 4\mathcal{H}u_k\rho_k + \rho_k \Phi_k=0\;,$$
and using the background formula from conservation equation of Eq.(\ref{eq:nonpertdensity}), we have
\begin{equation}\label{velocsubhoriz}
  k^2 \rho_0 u'_k = - k^2 \mathcal{H} u_k - k^2\Phi_k\;.
\end{equation}

Performing the definition of the ``contrast'' matter density $\delta_m \equiv \frac{\delta \rho}{\rho_0}$, and using Eqs.(\ref{tensorcompo00kspace2}), (\ref{denssubhoriz}) and (\ref{velocsubhoriz}), we obtain a relation with $\Phi_k$ and $\delta_m$ as
\begin{equation}\label{gravpotential}
  k^2 \Phi_k= -4\pi G_{eff} a^2\rho_0 \delta_m\;,
\end{equation}
where $G_{eff}$ is the effective Newtonian constant and is given by
\begin{equation}\label{eq:geff}
G_{eff}(a)=\frac{G}{1- 9\gamma_0 a^{2\beta_0}}\;,
\end{equation}
that results in a ``flat'' $G_{eff}$ in the sense that $G_{eff}$ is independent of \emph{k}-scale, like that of $f(T)$ models \cite{ Zheng11,ness11}.

From Eq.(\ref{eq:nonpertfriedtotal3}), a fluid analogy is provided by the effective Equation of State (EoS) for an ``extrinsic fluid'' parameter $w_{ext}$ by defining
\begin{equation}\label{eq:wext}
w_{ext}=-1+\frac{1}{3}\left(4-2\beta_0\right)\;.
\end{equation}
If we adopt the previous relation of the dimensionless parameter $\beta_0$ with the dark energy fluid parameter $w$ in such a way $\beta_0=2-\frac{3}{2}\left(1+w\right)$, or equivalently, $w=-1-\frac{1}{3}\left(2\beta_0-4\right)$, we obtain $w_{ext}=w$. Hence, we can write the dimensionless Hubble parameter $E(z)$ as
\begin{equation}\label{eq:dimenHub}
E^2(z)=\Omega_{m(0)}(1+z)^3+ \left(1-\Omega_{m(0)}\right)(1+z)^{3(1+w)}\;.
\end{equation}
which reproduces a quintessence or/phantom behaviour at background level depending on the value of $w\neq -1$. Hereon, the present model is denoted as $\beta$-model only to facilitate the referencing.
Consequently, $G_{eff}$ can be written as
\begin{equation}\label{eq:geff2}
G_{eff}(a)=\frac{G}{1-9\gamma_0 a^{1-3w}}\;,
\end{equation}
and using Eqs.(\ref{eq:extomega}) and (\ref{eq:gammaeq}), one obtains that $\gamma_0 $ can be rewritten as
\begin{equation}\label{eq:gammas}
\gamma_0= \frac{1}{2}\gamma_s(1-3w)(1-\Omega_{m(0)})\;.
\end{equation}
It is important to point out that due to Eq.(\ref{eq:wext}), which is an assumption that relates the geometrical approach of $\beta$-model to a physical cosmological parameter $w$, we also need to introduce a $\gamma_s $ parameter. The necessity is twofold: first, it is to maintain the reproducibility of GR/$\Lambda$CDM limit, i.e., when $\gamma_s\rightarrow 0$. Thus, even a small deviation from that limit value of $\gamma_s $  leads to a deviation from GR. Secondly, it is to sustain such a level of arbitrariness of $\gamma_0$  in Eq.(\ref{eq:gammaeq}) as a heritage from its extrinsic origin in order to stabilise the evolution of $G_{eff}$ due to the introduction of EoS for the ``extrinsic fluid''.

\begin{table*}
\scriptsize
\centering
\caption{Marginalised constrains on the cosmological parameters (mean values) from \texttt{GetDist} at $68\%$ limits of MCMC chains of each individual model. The $\chi_{tot}^2$ denotes the $\chi^2$ of the combined joint datasets. The $\chi^2$ of each individual dataset is also presented.}
  \begin{tabular}{|c|c|c|c|c|c|c|}
    \hline
    \multirow{2}{*}{Parameters} &
      \multicolumn{2}{c|}{\textbf{P18+R20}}&
      \multicolumn{2}{c|}{\textbf{P18+R20+BAO}}&
      \multicolumn{2}{c|}{\textbf{P18+BAO+R20+DESY1+Pantheon}}\\
                                          & $\Lambda$CDM & $\beta$-model & $\Lambda$CDM & $\beta$-model & $\Lambda$CDM & $\beta$-model \\
    \hline
   $H_0            $              & $68.45\pm 0.53$           & $69.02\pm 0.74$          & $68.43\pm 0.44$        &$68.87\pm 0.60$   & $68.62\pm 0.42$ & $69.05\pm 0.56$\\

   $\Omega_\mathrm{m}         $              & $0.3007\pm 0.0070   $   & $0.2955\pm 0.0085$      & $0.3009\pm 0.0057$     & $0.2971\pm 0.0068$    & $0.2984\pm 0.0053$& $0.2950\pm 0.0063$ \\

   {\boldmath$\Omega_\mathrm{b} h^2$}        & $0.02255\pm 0.00014$    &  $0.02258\pm 0.00014$                    & $0.02255\pm 0.00013$   & $0.02257^{+0.00015}_{-0.00013}$     & $0.02258\pm 0.00013$& $0.02257\pm 0.00013$ \\

   {\boldmath$\Omega_\mathrm{c} h^2$}        & $0.1177\pm 0.0012$   & $0.1175\pm 0.0012$   & $0.11770\pm 0.00098$     & $0.11769\pm 0.00092$    & $0.11728\pm 0.00092$& $0.11741\pm 0.00088$\\

   {\boldmath$\tau_\mathrm{reio}$}           & $0.0581\pm 0.0083$   & $0.0553^{+0.013}_{-0.0087}$             & $0.0579\pm 0.0082$   & $0.0537^{+0.014}_{-0.0087}$  & $0.0580^{+0.0073}_{-0.0088}$& $0.0579^{+0.0071}_{-0.0084}$\\

   {\boldmath$w              $}              & $-1$                             & $< -1.59$                         & $-1$                     & $<-1.54$      & $-1$& $< -1.8$\\

   {\boldmath$\log(10^{10} A_\mathrm{s})$}   & $3.047\pm 0.017$   & $3.043\pm 0.20$   & $3.047\pm 0.017$       & $3.041\pm 0.020$  &$3.046^{+0.015}_{-0.018}  $    & $3.046^{+0.015}_{-0.017}      $\\

   {\boldmath$n_\mathrm{s}   $}              & $0.9706\pm 0.0040$   & $0.9706\pm 0.0040 $   &$0.9705\pm 0.0036$      &$0.9701\pm 0.0034 $      &$0.9712\pm 0.0038 $ & $0.9707\pm 0.0037 $\\

   {\boldmath$100\theta_\mathrm{s}$}         & $1.04209\pm 0.00029            $   & $1.04212\pm 0.00026 $   & $1.04209\pm 0.00029$   & $1.04212\pm 0.00026 $      &$1.0421\pm 0.00027 $ & $1.0421\pm 0.00030 $\\

   $A_\mathrm{s} (\times 10^{9})$           & $2.105\pm 0.036$   & $2.098\pm 0.42$         & $2.105\pm 0.036$ & $2.093\pm 0.042$ & $2.103^{+0.031}_{-0.038}$& $2.103^{+0.031}_{-0.037}$ \\

   $\sigma_8       $              & $0.8051\pm 0.0078$        & $0.8069\pm 0.0084$ & $0.8052\pm 0.0076$     & $0.8061\pm 0.0086$    & $0.8035^{+0.0067}_{-0.0075}$  & $0.8081\pm 0.0074$ \\

   $S_8\equiv \sigma_8 \left(\Omega_\mathrm{m}/0.3\right)^{0.5}$        & $0.806\pm 0.015    $     & $0.801^{+0.012}_{-0.015}$                  & $0.806\pm 0.013$   & $0.802^{+0.010}_{-0.013}$& $0.8014\pm 0.012$ & $0.801\pm 0.012$ \\

    $\beta_0 $             & $---$       & $> 2.885$    & $---$    & $> 2.81$            & $---$                    & $> 3.2$ \\

    $\gamma_0 $        & $---$       & $0.00204\pm 2.46\times 10^{-5}$    & $---$   & $0.00199\pm 1.91\times 10^{-5}$            & $---$                   & $0.00226\pm 2.02\times 10^{-5}$ \\

   $\chi_{tot}^2                    $        & $2796.5\pm 5.8            $     & $2795.4\pm 6.8 $   & $2800.0\pm 5.9 $  & $2799.7\pm 7.0 $ & $3914.8\pm 6.7 $& $3912.7\pm 6.6 $ \\

   $\chi^2_\mathrm{planck\ 2018\ lowl.TT}$   & $22.48\pm 0.77$ & $22.43\pm 0.74$  & $22.49\pm 0.72             $   & $22.49\pm 0.67$ & $22.38\pm 0.75 $ & $22.45\pm 0.73             $  \\

   $\chi^2_\mathrm{planck\ 2018\ lowl.EE}$   & $397.7\pm 2.5$ & $397.9 \pm 2.4$  & $397.7\pm 2.5              $   & $397.8\pm 2.3              $ & $397.6\pm 2.5$ & $397.6\pm 2.4              $ \\

   $\chi^2_\mathrm{planck\ 2018\ high\emph{l}.TTTEEE }$ & $2362.8\pm 6.9$ & $2364.4\pm 6.7$        & $2362.5\pm 6.6 $        & $2364.0\pm 6.3 $ & $2364.0\pm 6.7 $ & $2363.4\pm 6.4 $  \\

   $\chi^2_\mathrm{bao.sdss\ dr12\ consensus\ bao}$  & $---$  & $---$  & $3.76\pm 0.52$ & $3.87\pm 0.59$  & $3.81\pm 0.6$ &$3.97\pm 0.66$\\

   $\chi^2_\mathrm{des\ y1.clustering}$              & $---$       & $---$ & $---$ & $---$  & $79.7\pm 3.3 $ & $79.8\pm 3.3 $ \\

   $\chi^2_\mathrm{H0.riess2020}$                    & $13.5\pm 3.0$       & $10.7\pm 3.6  $ & $13.6\pm 2.5$ & $11.3\pm 3.0$  & $12.5\pm 2.3 $& $10.4\pm 2.8 $ \\

   $\chi^2_\mathrm{SnIa.Pantheon}$                   & $---$                & $---$          & $---$ & $---$  & $1034.794\pm 0.083 $& $1035.20\pm 0.54 $ \\
  \hline
  \end{tabular}\label{table:2}
\end{table*}

In order to analyse the evolution of matter density, we first take the conformal time derivative of Eq.(\ref{denssubhoriz}). Then, writing Eq.(\ref{velocsubhoriz}) in terms of $\delta_m$ and using Eq.(\ref{gravpotential}), we obtain the equation of evolution of the contrast matter density $\delta_m(\eta)$ in conformal longitudinal Newtonian frame
\begin{equation}\label{contrastdensityconformal}
\delta''_m + \mathcal{H}\delta'_m - 4\pi G_{eff} a^2\rho_0 \delta_m=0\;.
\end{equation}

To express Eq.(\ref{contrastdensityconformal}) in terms of the physical time, we use the notation $\dot{\delta}_m\equiv \frac{\delta'_m}{a}$ and obtain the useful relation $\delta''_m= a^2 \ddot{\delta}_m + a^2H \dot{\delta}_m$. We are leading to
\begin{equation}\label{contrastdensityphysical}
\ddot{\delta}_m(t)+ 2H\dot{\delta}_m(t) - 4\pi G_{eff}\rho_0 \delta_m(t)=0\;.
\end{equation}
Thus, we obtain an alternative way to express the former equation in terms of the expansion factor $a(t)$. We use the notation $\frac{d\delta_m(a)}{da}= \delta^{\circ}_m(a)$ and $\frac{d^2\delta_m(a)}{da^2}= \delta^{\circ\circ}_m(a)$, and obtain useful relations
$\delta^{\circ}_m(a)= \frac{1}{\dot{a}}\dot{\delta}_m(t)\;$ and $\delta^{\circ\circ}_m(a)= \frac{1}{\dot{a}^2}\ddot{\delta}_m(t)$. Hence, the contrast matter density $\delta_m(a)$ is governed by the equation
\begin{equation}
\delta^{\circ\circ}_m(a)+\left(\frac{3}{a}+\frac{H^{\circ}(a)}{H(a)}\right)\delta^{\circ}_m(a)-\frac{3\Omega_{m0}G_{eff}/G}{2(H^2(a)/H_0^2)}\delta_m(a)=0\;.\label{eq:constrastmatter}
\end{equation}
which solutions are possible only numerically. For instance, in the context of GR, where $G_{eff}=G$ turns $\delta_m(a)$ independent of the scale \emph{k}, with the fluid parameter $w$, one has the following solution
\begin{equation}\label{eq:constrastsolut}
\delta(a) = a . 2F_1\left(-\frac{1}{3w},\frac{1}{2}-\frac{1}{2w} ; 1-\frac{5}{6w} ; a^{-3w}(1-\Omega_m^{-1})\right)\;
\end{equation}
where $2F_1(a,b;c;z)$ is a hypergeometric function. Clearly, the presence of $G_{eff}$ function in Eq.(\ref{eq:constrastmatter}) marks the departure of the $\beta$-model from any analogy with $\Lambda$CDM ($w=-1$) at perturbation level.

The form of the effective Newtonian constant is given by Eq.(\ref{eq:geff}) as a result of the linear Nash-Greene fluctuations of the metric and the induced extrinsic curvature as shown in Eqs.(\ref{eq:g}) and (\ref{eq:k1}).

\section{Analysis on tensions of Hubble parameter and growth amplitude factor}
\begin{table}
\centering
\caption{Flat priors on the cosmological parameters used in our MCMC analysis. }
\begin{tabular} {cc}
\hline
 Parameter                                       & Priors       \\
\hline
{\boldmath$\Omega_\mathrm{b} h^2$}               & [0.001,0.3]  \\

{\boldmath$\Omega_\mathrm{cdm} h^2$}             & [0.001,0.99]  \\

{\boldmath$\tau_\mathrm{reio}$}                  & [0.01,0.8] \\

{\boldmath$w              $}                     & [-3,0] \\

{\boldmath$n_\mathrm{s}   $}                     & [0.8,1.2] \\

{\boldmath$\ln(10^{10}A_s)  $}                   & [1.61,3.91] \\

{\boldmath$100\theta_{MC}    $}                  & [0.5,10]  \\
\hline
\end{tabular}\label{table:1}
\end{table}
In this section, we focus on the growth amplitude factor $S_8$, Hubble constant $H_0$ and matter density $\Omega_m$, since they are the main actors of the discrepancy problem in cosmological probes. We basically compare the results of our model with the minimal flat $\Lambda$CDM model.  For the numerical implementation, we wrote a code using \texttt{Cobaya} \cite{cobaya1,cobaya2} sampler and the module \texttt{Classy} to include the cosmological theory code \texttt{CLASS} \cite{class1,class2,class3} by means of a modification \texttt{EFCLASS}\cite{radou2} to better define the perturbation equations of the model. Concerning the $\Lambda$CDM model, the \texttt{CLASS} code was kept intact and all the related chains ran using the standard \texttt{Cobaya} vanilla code. In order to keep the analysis on the sub-horizon linear scale, we set in the  \texttt{EFCLASS} code the minimum value of expansion parameter as $a_{min}=0.001$.

The joint analysis is made by using of data on the evolution of background parameter $H(z)$ and $\Omega_m$ distributions from Planck likelihood code \cite{planck2018} with 2810 points, SNIa Pantheon data \cite{Pantheon} with 1048 points, the galaxy clustering and weak lensing from DES Y1 \cite{DES} with 90 points, the BAO SDSS DR12 ``consensus'' galaxy sample \cite{DR12} with 6 points from redshift-space distortions (RSD) to compute gro-wth data, $H(z)$ and BAO.  As of writing, it is important to point out that we opt not to include other RSD measurements due to (still) unsolved issue in BAO+RSD likelihood\footnote{\url{https://github.com/CobayaSampler/cobaya/issues/44}.} once f$\sigma_8$ is not implemented  so far in CLASS and produces an error when calculating by \texttt{Cobaya}. From Planck 2018 data \cite{planck2018}, we consider CMB temperature and polarisation angular power spectra (high-\emph{l}.plik.TTTEEE + low-\emph{l} EE polarisation+ low-\emph{l} TT temperature) quoted simply as P18. As local measurements on Hubble constant $H_0$, we use Riess et al. 2020 \cite{riess20} data, hereon R20, from Hubble Space Telescope (HST) photometry and Gaia EDR3 parallaxes with $H_0 = 73.2\pm 1.3$km$s^{-1}$.Mpc$^{-1}$ with 20 datapoints. For model comparison, we adopt the following combination of joint dataset as P18+R20, P18+R20+BAO and P18+R20+BAO+\hspace{0.2cm}+DESY1+Pantheon.

The MCMC chains are analysed by using \texttt{GetDist} \cite{getdist} in which was also used to make the contour plots. We sample the posterior distributions of the MCMC chains by means of Metropolis–Hastings algorithm \cite{mcmc1,mcmc2} in \texttt{Cobaya}. The parallel runs were stopped by applying the Gelman-Rubin convergence criterion \cite{gelman} $R-1 <0.02$ and the first 30$\%$ of chains were discarded as burn-in for each sample of joint data. We summarise the results of MCMC analyses in Table (\ref{table:2}) with the mean marginalised posterior values for the parameters. We adopt baseline priors as shown in Table (\ref{table:1}): the baryon density is given by $\Omega_\mathrm{b} h^2$, $\Omega_\mathrm{cdm}h^2$ represents CDM density, the reonisation optical depth $\tau_\mathrm{reio}$, dark fluid parameter $w$, scalar spectral index $n_\mathrm{s} $, the amplitude of primordial fluctuations $\ln(10^{10}A_s)$ and the angular size of the first CMB acoustic peak $100\theta_{MC}$. In the case of $\Lambda$CDM, the dark fluid parameter is fixed as $w=-1$.

\begin{figure*}[htp]
  \centering
  \mbox{
  \includegraphics[width=9.15cm, height=10cm]{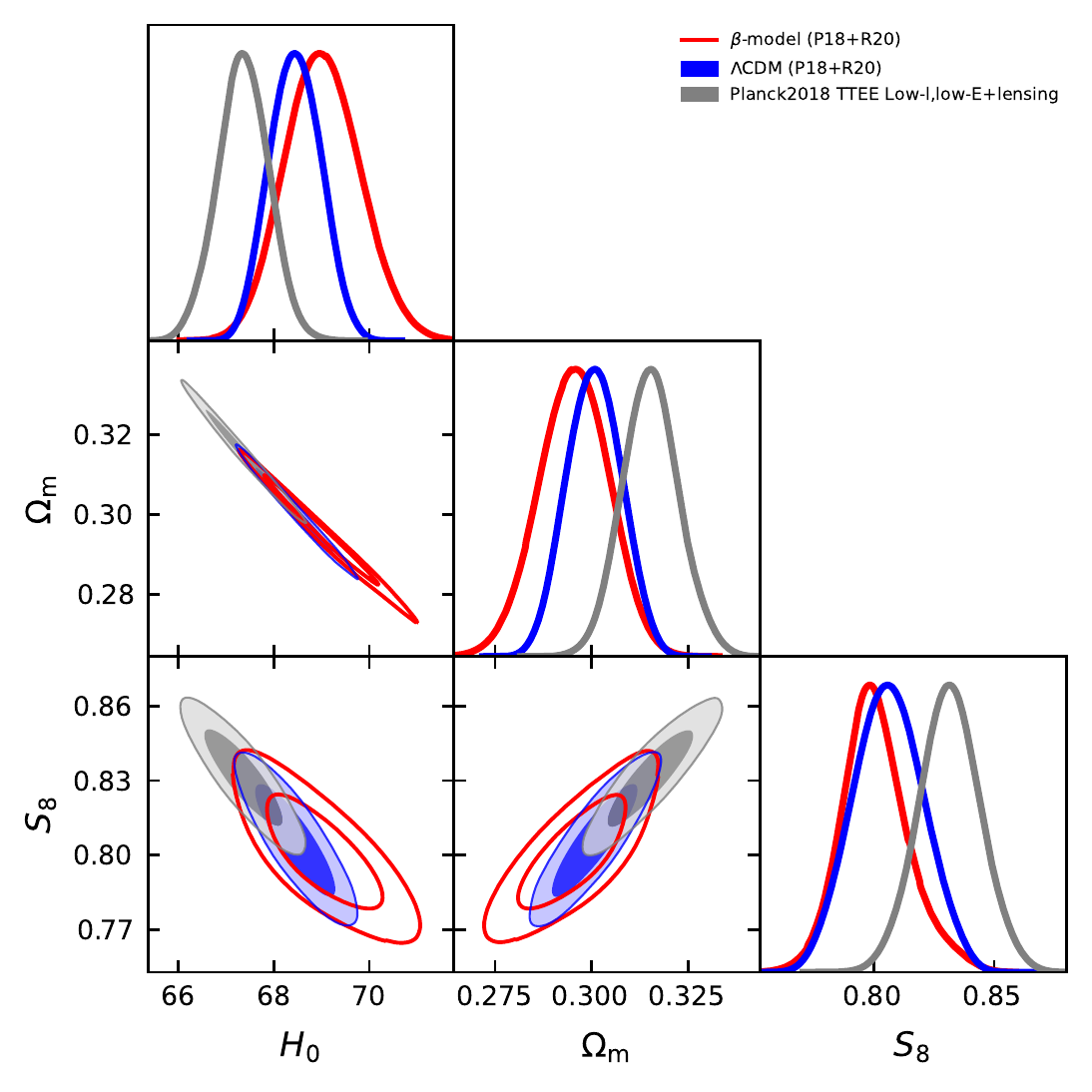}
  \includegraphics[width=9.15cm, height=10cm]{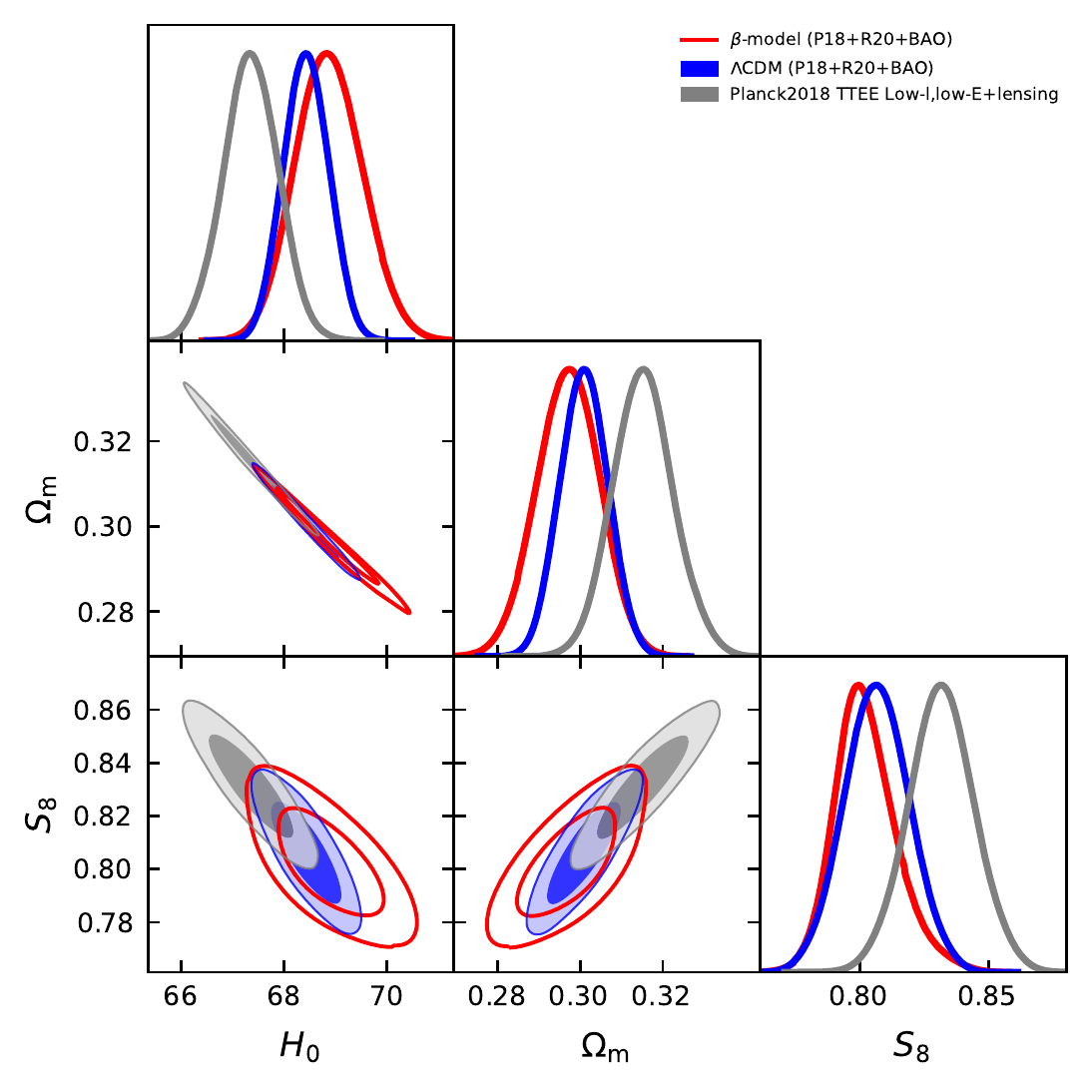}}
  \mbox{
  \includegraphics[width=10.2cm, height=11cm]{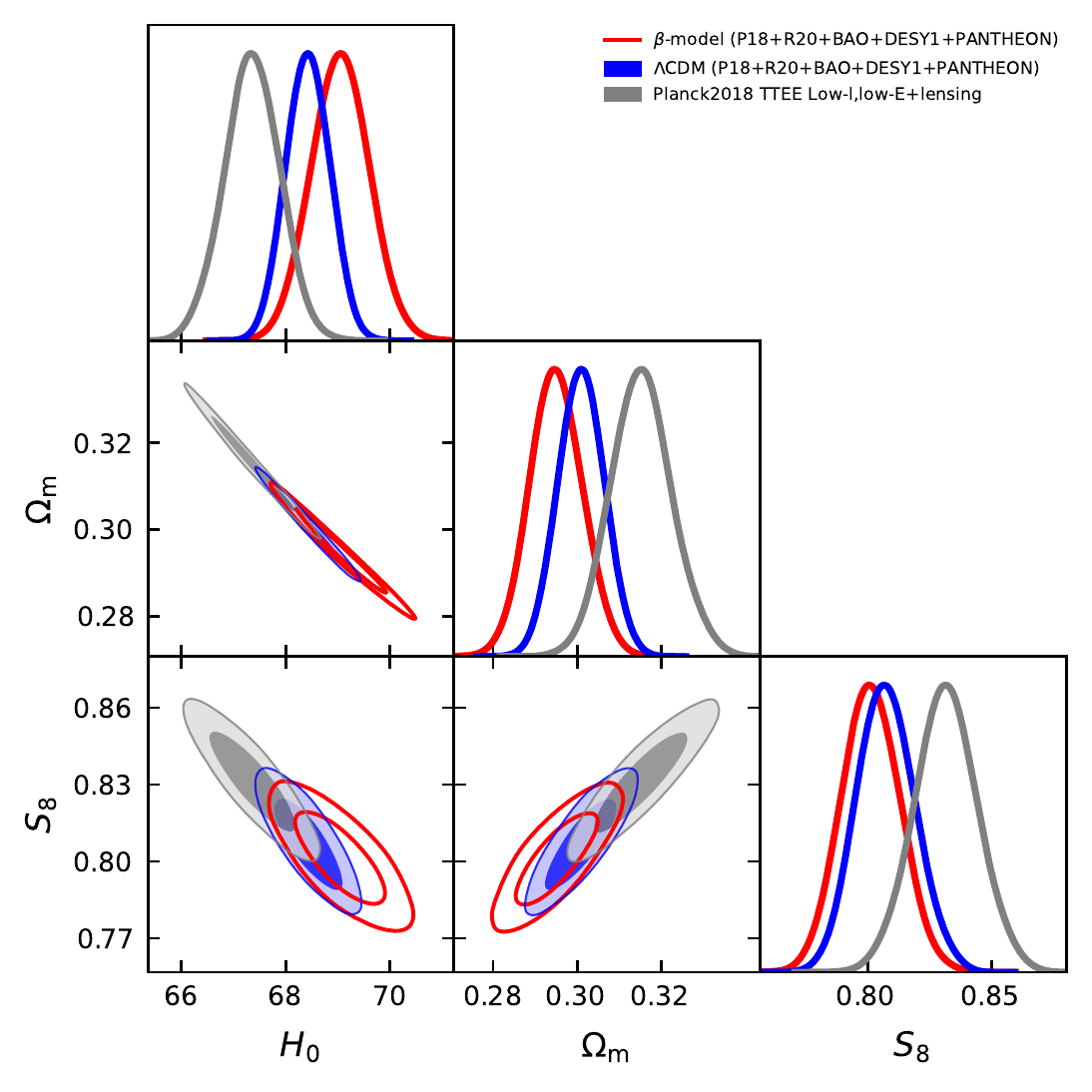}}
  \caption{The one-dimensional marginalised posterior distributions and two-dimensional contour plots with $68.4\%$ and $95.7\%$ C.L. of cosmological parameters ($\Omega_m, S_8,H_0$). Red and Blue colours indicate the $\beta$-model and $\Lambda$CDM, respectively. The grey colour indicates the Planck 2018 high-\emph{l} temperature TTTEEE+low-\emph{l}+low-E+lensing data. (For interpretation of the references to colour in this figure legend, the reader is referred to the web version of this article.)}\label{fig:fig1}
\end{figure*}
\begin{figure}[htp]
\centering
\includegraphics[width=8.15cm, height=9.5cm]{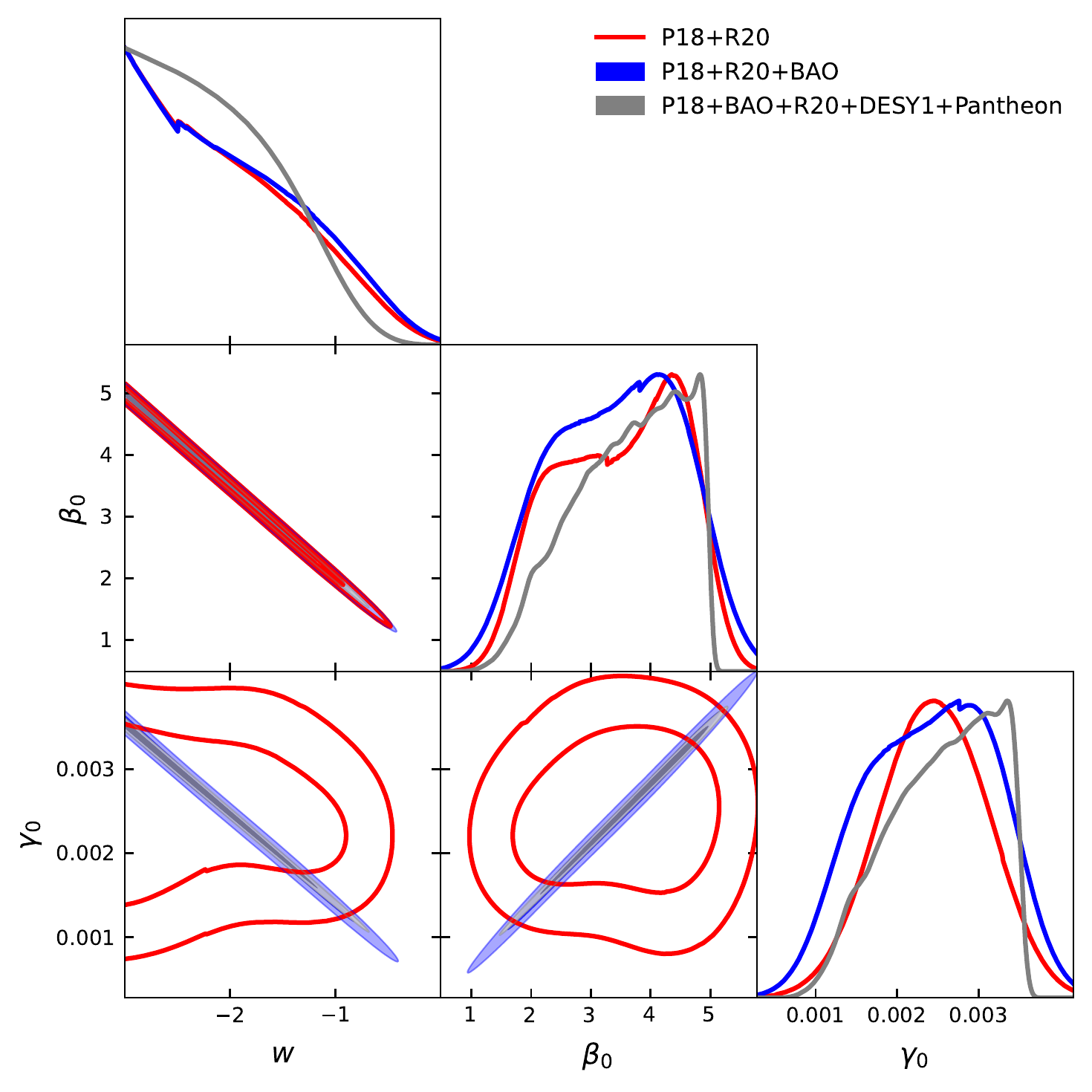}\quad
  \caption{Red, blue and grey colours indicate the resulting contours for the  $\beta_0$, $\gamma_0$ and $w$ parameters at $68.4\%$ and $95.7\%$ C.L. of P18+R20+BAO, P18+R20+BAO+DESY1+Pantheon joint data, respectively. (For interpretation of the references to colour in this figure legend, the reader is referred to the web version of this article.)}\label{fig:fig2}
\end{figure}

In order to check the modification of the value of the gravitational constant during Big Bang nucleosynthesis (BBN) epoch as $z_{\rm BBN} \sim 10^9$, we calculate the BBN speed-up factor \cite{uzan,sola}. At the BBN epoch the bound is about $|\frac{\Delta H^2}{H^2_{\Lambda CDM}}|< 10 \%$. From the values of MCMC chains, we obtain the BBN speed-up factor between $\Lambda$CDM and the $\beta$-model is roughly 1$\%$, 0.7$\%$ and 0.6$\%$ from the joint datasets P18+R20, P18+\hspace{0.2cm}+R20+BAO and P18+R20+BAO+DESY1+Pantheon, respectively. In all cases, they largely satisfy the bounds on BBN speed-up factor. Concerning $G_{eff}$ in Eq.(\ref{eq:geff}), we also need to check if it obeys BBN constraints. To do so, we have rewritten Eq.(\ref{eq:geff2}) with its right-shifted parent function as
\begin{equation}\label{eq:geff3}
G_{eff}(a)=\frac{G_N}{1-9\gamma_0 (a-1)^{1-3w}}\;.
\end{equation}
This parent function only changes from the growth pattern of the original function in Eq.(\ref{eq:geff2}) to a decaying behaviour.  To guarantee the positivity of $G_{eff}>0$, then $\gamma_s$ must comply with
\begin{equation}\label{eq:gammas2}
\gamma_s< \frac{0.111}{(1-3w)(1-\Omega_{m(0)})}\;.
\end{equation}
For  P18+R20, we find $\gamma_s< 0.0273\pm 0.0003$, P18+R20+BAO with $\gamma_s< 0.0281\pm 0.0003$ and P18+R20+BAO+DESY1+\hspace{0.5cm}+Pantheon with $\gamma_s< 0.0244\pm 0.0002$. We notice that since $w$ and $\Omega_{m(0)}$ have bounded values, fixing $\gamma_s\leq 1\times10^{-3}$ will suffice for all cases and it attends the condition of Eq.(\ref{eq:gammas2}) and  $G_{eff}$ constraints. Hence, $\gamma_s$ is a data independent parameter. The form of Eq.(\ref{eq:geff}) complies with the constraint $\frac{G_{eff}}{G_N}\rfloor_{a=0}=1$, $\frac{G_{eff}}{G_N}\rfloor_{a=1}=1$ expected for both BBN and solar scales for any $w<0$. We obtain $\frac{G_{eff}}{G}\rfloor_{a=0}\sim 1.02 $ and $\frac{G_{eff}}{G}\rfloor_{a=1}=1$ , which obeys BBN constraints $|G_{eff}/G -1|\leq 0.2$ \cite{craig}. Regardless of the value of $\gamma_0$, we also obtain the derivative $\frac{d G_{eff}}{da}/G\rfloor_{a=1}=0$, that obeys the constraint $\frac{d G_{eff}}{da}/G\rfloor_{a=1}\simeq 0$ \cite{nesseris2007}. For early times, considering that the BBN constraint is not so stringent \cite{Nesseris2017}, we have $\frac{d G_{eff}}{da}/G\rfloor_{a=0}> 0$. For completeness purposes, in Fig.(\ref{fig:fig2}) we also present the contours on the constrained $\beta_0$ and $\gamma_0$ parameters in contrast with the fluid parameter $w$.  The posterior probability of $\beta_0$ and $\gamma_0$  appears to be multimodal.

\begin{table*}[htp]
\caption{A summary  of the obtained values of AIC, MBIC and HQC for the studied models. $\Lambda$CDM is adopted as a reference model.}\label{tab:complaic}
\begin{tabular}{p{0.23
\linewidth}p{0.08\linewidth}p{0.06\linewidth}p{0.07\linewidth}p{0.07\linewidth}p{0.07\linewidth}p{0.07\linewidth}p{0.07\linewidth}p{0.07\linewidth}}
\hline
Model                       &$AIC$ & $\Delta AIC$ & \textbf{Evidence }      &$MBIC$ & $\Delta MBIC$      &$HQC$     & $\Delta HQC$    & \textbf{Evidence} \\
\hline
$\Lambda$CDM (full)         & 3922.40  & $\;\;$0  & null           & 3940.6   & $\;\;$0           & 3931.72  & $\;\;$0          & null     \\
$\beta$-model(full)         & 3922.72  & 0.1      & weak           & 3944.95  & 4.34             & 3933.85  & 2.13       & Positive  \\
$\Lambda$CDM (P18+R20+BAO)  & 2808.01  & $\;\;$0  & null           & 2824.45  & $\;\;$0           & 2816.59  & $\;\;$0          & null     \\
$\beta$-model(P18+R20+BAO)  & 2809.72  & 1.71     & weak           & 2830.26  & 5.81            & 2820.43  & 3.85       & Positive  \\
$\Lambda$CDM (P18+R20)      & 2805.51  & $\;\;$0  & null           & 2820.94  & $\;\;$0           & 2813.08  & $\;\;$0          & null     \\
$\beta$-model(P18+R20)      & 2804.51  & 0.9      & weak           & 2025.95  & 5.01             & 2816.13  & 3.04       & Positive  \\
\hline
\end{tabular}
\end{table*}
In this paper, we use as a reference for model comparison three information criteria (IC) classifiers to estimate the strength of tension between the data fitting and particular models using maximum likelihood estimation. To sum up, the model with higher IC tends to aggravate the tension when more (free) parameters are allowed, and the simpler model is normally preferable rather than the complex one. In cosmology, this situation must be seen with caution once considering only the maximum likelihood analysis it may induce false positives (or negatives) and choose an ultimate reliable model turns a non trivial task to decide between the model simplicity over complexity \cite{liddle2004, trotta2007, trotta2011, nesseris2013, nesseris2016}.

Adopting the data as being Gaussian, we use the Akaike criterion (AIC) \cite{Akaike} for small samples sizes \cite{liddle, Sugiura} such as
\begin{equation}\label{eq:aic}
AIC= \chi^2_{bf} + 2k\frac{2k(k+1)}{N-k-1}\;,
\end{equation}
where $\chi^2_{bf}$ is the best fit $\chi^2$ of the model, $k$ represents the number of the uncorrelated (free) parameters and $N$ is the number of the data points in a dataset. The difference $|\Delta AIC|= AIC_{model\;(2)}-AIC_{model\;(1)}$ represents the Jeffreys' scale \cite{jeff} that qualitatively classifies the intensity of tension between two (competing) models. Higher values for the difference $|\Delta AIC|$ indicates more tension between the models and thus more statistically uncorrelated they are. Jeffreys' scale stands for $|\Delta AIC\leq 2|$ in which the tension is weak and the models are statistically consistent with a considerable level of empirical support. For $4< \Delta AIC < 7 $ indicates a positive tension against the competing model with a higher value of AIC. For $|\Delta AIC \geq 10|$ indicates a strong evidence against the model with a higher AIC. The other two IC classifiers tend to impose a major penalty on model complexities. In order to correct the original BIC \cite{bic} that does not get rid of change-point processes (which is particulary important in MCMC processes), the Modified Bayesian Information Criterion (MBIC) \cite{mbic}, is given by the formulae
\begin{equation}\label{eq:mbic}
MBIC = \chi^2_{bf} + k (\ln[N]-\ln[2\pi])\;.
\end{equation}
From the Jeffreys' scale, as in the AIC case, higher values for $|\Delta MBIC|$ denotes more tension between competing models and more statistically uncorrelated they are. Unlike AIC, MBIC requires a better detailed Jeffreys' scale for a qualitative analysis of model comparison. For instance, $|\Delta MBIC\leq 2|$ indicates that is not worth more than a bare mention (about tension) and the models are statistically consistent with a considerable level of empirical support. For $2< \Delta MBIC \leq 6 $ indicates a positive tension against the model with a higher value of BIC. For $6 < \Delta MBIC \leq 10 $, it defines a strong evidence against the model with a higher MBIC value and a very strong evidence against the model with a higher BIC value happens when $\Delta MBIC > 10$.

Moreover, the last classifier relies on Hannan–Quinn information Criterion (HQC) \cite{hqc} as an alternative to AIC and MBIC based on the law of the iterated logarithm. It penalises the complex model with an $(\ln \ln N)$ factor by
\begin{equation}\label{eq:hqc}
HQC = \chi^2_{bf} + 2 k \ln[\ln[N]]\;.
\end{equation}
Concerning Jeffreys' scale, we use the same arrange like that of MBIC. As about to be seen, MBIC penalises additional parameters of a model much heavily than AIC and HQC. The results are presented in Table (\ref{tab:complaic}) in which we compare the models in the light of their response from different datasets. The ``full'' data stands for P18+R20+BAO+DESY1\\+Pantheon. As a result, for all cases, $\Delta AIC$ indicates a weak tension between the models and a positive support by the data for $\beta$-model, being $\Lambda$CDM adopted as a reference model. Both $\Delta MBIC$ and $\Delta HQC$ indicate a positive tension. An interesting situation rises from the fact that the inclusion of the BAO data (except for the case of ``full'' data), concerning the $\beta$-model, signs a direction towards to the low limit ($\Delta HQC$) and the higher limit ($\Delta MBIC$) of Jeffreys' scale in the positive tension classification. This merits further research since the inclusion of more BAO data may aggravate or not the IC values.

As a matter of comparison, we use the full chains of Planck 2018 high-\emph{l} temperature TTEE low-\emph{l}+low-E+lensing \cite{planck2018} \footnote{Available at \url{https://pla.esac.esa.int/pla/$\#$home}} to reproduce the grey contours in Fig.(\ref{fig:fig1}) with different joint data. In the plots, our model ($\beta$-model) is represented by the red colour and $\Lambda$CDM by the blue colour (the reader is referred to the web version of this article). The left and right upper panels present a similar pattern for P18+R20 and P18+R20+BAO. Concerning the Hubble tension, when compared with R20 of $H_0 = 73.2\pm 1.3$km$s^{-1}$.Mpc$^{-1}$, we obtain at $68.4\%$ $\sim 2.67\sigma$ and $\sim 2.89\sigma$, which is lower than the values obtained for $\Lambda$CDM, with $\sim 3.25\sigma$ and $\sim 3.34\sigma$, respectively. We point out the particular influence of BAO which tends to worsen the discrepancy on $H_0$, which was already observed in the IC analysis. This reinforces the problem with BAO data, which is model dependent. In the lower panel, when considering P18+R20+BAO+DESY1+Pantheon, we obtain $\sim 2.81\sigma$, which is lower than the one observed with P18+R20+BAO, and better than the value obtained with $\Lambda$CDM that present a discrepancy around $3.22\sigma$. In all cases for $\beta$-model, we obtain a negative correlation between $\Omega_m$ and $H_0$, and the dark fluid parameter $w$ prefers a phantom EoS in accordance with Planck data. At $95.7\%$ C.L., we obtain fluid parameter $w<-0.978$, $(<-0.982)$ and $(<-1.02)$; the values for $H_0$ as $69\pm 1.6$, $68.9\pm 1.3$ and $69\pm 1.1$ that give $\sim 1.95\sigma$, $\sim 2.24\sigma$ and $\sim 2.36\sigma$ for P18+R20, P18+R20+BAO and P18+R20+BAO+DESY1+Pantheon, when compared with R20, respectively. Our resulting mean values on $H_0$ are close to the observed one for nearby galaxies with a red giant branch (TRGB) with $H_0 = 69.8\pm 1.1$km$s^{-1}$.Mpc$^{-1}$ \cite{freed20}.
\begin{figure}[htp]
\centering
\includegraphics[width=8.6cm, height=7.2cm]{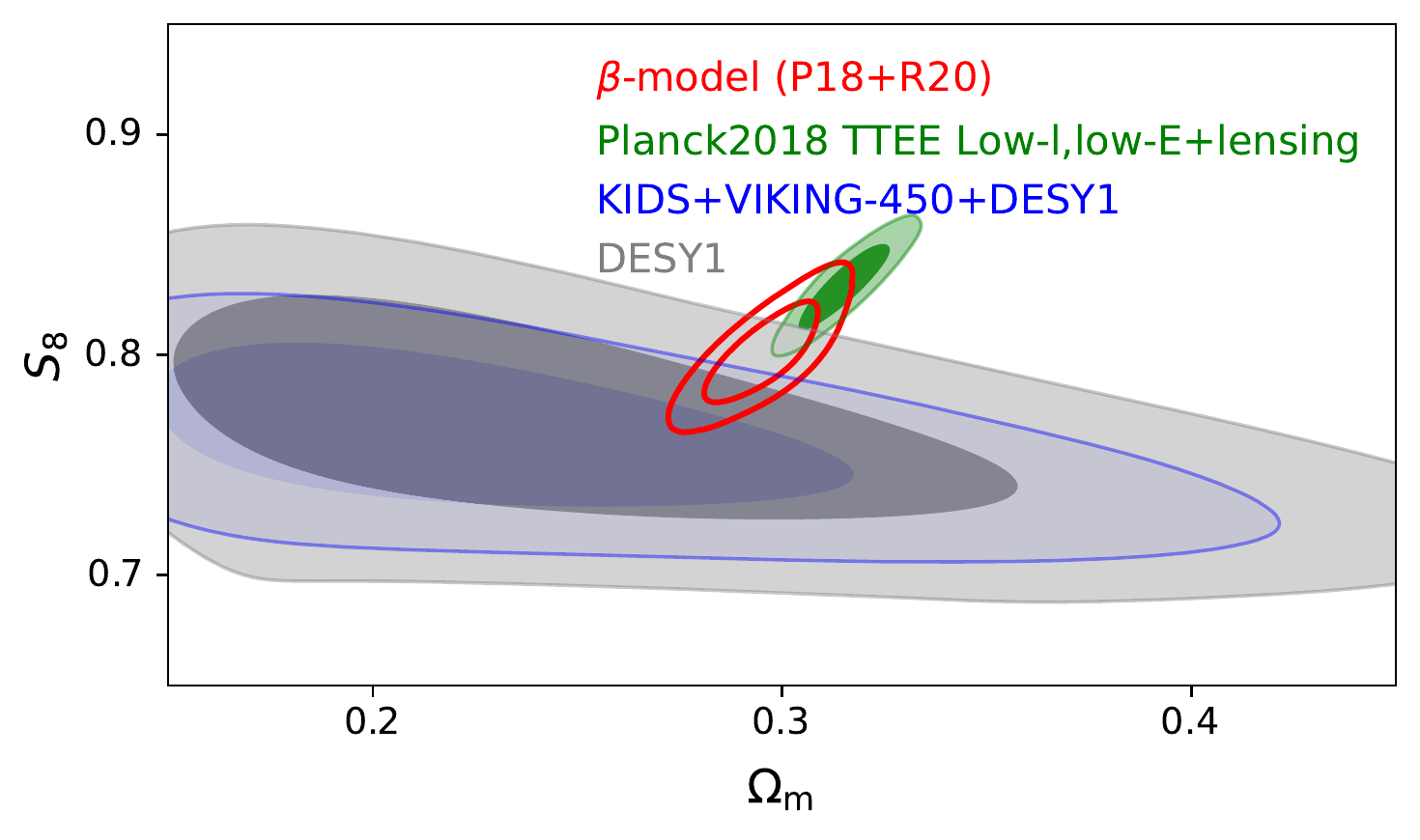}\quad
  \caption{The two-dimensional contour plots at $68.4\%$ and $95.7\%$ C.L. in the plane ($S_8-\Omega_m$). Red, green and blue colours indicate the $\beta$-model for P18+R20 joint data, Planck 2018 high-\emph{l} temperature TTTEEE+low-\emph{l}+low-E+lensing data and KIDS+VIKING-450 in combination with DESY1 data, respectively. The grey colour indicates DESY1 data. (For interpretation of the references to colour in this figure legend, the reader is referred to the web version of this article.)}\label{fig:fig3}
\end{figure}

Concerning the amplitude discrepancy on $S_8$ parameter, we compare our results with DESY1\cite{DES} shear results adopting the $S_8$ parameter that has the values of $S_8=0.773^{+0.026}_{-0.020}$ (with $\Lambda$CDM background) at $68.4\%$CL.  Hence, from P18+R20, P18+R20+BAO and P18+R20+BAO+DESY1+Pantheon, we obtain $\sim 0.93\sigma$, $\sim 0.99\sigma$ and $\sim 0.978\sigma$, which reveal a reducing of tension as compared with those values obtained for $\Lambda$CDM with $\sim 1.1\sigma$ and $\sim 1.14\sigma$ for the two first joint data. On the other hand, we obtained for P18+R20+BAO+\hspace{0.2cm}+DESY1+Pantheon that $\Lambda$CDM tension is around $\sim 0.98\sigma$ maintaining an agreement with $\beta$-model. In all cases, $\beta$-model indicates a good agrement with the DESY1 probe and thus an alleviation of the $S_8$ tension. Again, BAO tends to worsen sightly the discrepancy on $S_8$.

In Fig.(\ref{fig:fig3}), it is shown the constraints on $\Omega_m$ and $S_8$ for the joint data in contrast with DESY1 alone with $\Lambda$CDM background.  It is also shown a comparison with KIDS plus VISTA-Kilo Degree Survey Infrared Galaxy Survey (KIDS+\\VIKING-450) and DESY1 combined \cite{Joudaki} with $S_8=0.762^{+0.025}_{-0.024}$ that has a discrepancy with Planck $\Lambda$CDM about $2.5\sigma$. For the studied dataset, we obtained P18+R20, P18+R20+BAO and P18+R20+BAO+DESY1+Pantheon, we obtain $\sim 1.34\sigma$, $\sim 1.42\sigma$ and $\sim 1.41\sigma$, respectively.

For Kids1000 \cite{KiDs1000} that has the value of $S_8=0.766^{+0.020}_{-0.014}$ at $68.4\%$CL, we obtained a slight better result as compared with KIDS+VIKING-450. We find values above $1\sigma$, that is, $1.4\sigma$ for P18+R20, $\sim 1.51\sigma$ for both P18+R20+BAO and P18+R20+BAO+DESY1+Pantheon scenarios indicating a soft tension between the probes.

\section{Remarks and prospects}
In this paper, we have studied cosmic perturbations of matter in a search of understanding if the contribution of the extrinsic curvature to complement Einstein's gravity is rather than semantics and relies on physical reality. It may pinpoint a renewed consideration of the concept of curvature, which has been regarded a paramount element of contemporary physics, as a fundamental physical agent itself. We work at two different (but correlated phases). First, we present the mathematical framework based on the embedding of Riemannian geometries in order to obtain the main expression of the embedded space-time deformations summarised in the extrinsic curvature perturbation $\delta k_{\mu\nu}$ and on how its projection is possible onto the embedded space-time by means of the Nash flow. We have shown that for dynamical embeddings, the perturbation coordinate $\mathbf{y}$, that accesses the ambient space, does not appear in the four dimensional confined line element, unlike those of rigid embedding models that need additional arguments to generate perturbations, such as the bulk equations and/or  any additional principle like Israel-Lanczos condition that replaces the dynamics of extrinsic curvature. Rather, we explore the bare effect of the dynamics of extrinsic geometry showing that the effects of the orthogonal perturbations of the background geometry are transferred by $\delta k_{\mu\nu}$ into the perturbed embedded space-time. The integrability conditions (Gauss, Codazzi and Ricci equations)  guarantee that from the dynamics of the four-dimensional space-time $V_4$ we obtain information of the bulk space and vice-versa since Riemann curvature of bulk space acts as a reference for the Riemann curvature of the embedded space-time. Then, only the induced perturbed four-dimensional equations will suffice to analyse the physical folding and consequences. From the linear Nash-Greene perturbations of the metric, we have shown how to transpose the initial process in the background metric of the embedding of geometries to trigger the perturbations by the Lie transport. Secondly, in order to construct a viable physical model, we have focused on the embedded space-time with the obtainment of the induced field equations and on the related the perturbed field equations where the cosmological perturbation theory is applied.

An interesting fact resides that in five dimensions the gravitational tensor equation is indeed a perturbed equation, once the perturbation of the Codazzi equation does not propagate cosmological perturbations being hampered by linear Nash's fluctuations. On the other hand, this presented landscape is dramatically different in $dim \geq 6$ with appearance of new geometric objects, such as the third fundamental form $A_{\mu\nu a}$ that is associated to gauge fields, which should be a theme of future research. We also calculated the longitudinal Newtonian gauge of this framework in the simplest case that the gravitational potentials coincide $\Psi=\Phi$. Moreover, we have obtained in the subhorizon scale the contrast matter density ignited by the embedding equations. The finding of the matter overdensity equation $\delta_m$ is a paramount quantity for latter studies to identify any signature of modifications of gravity due to cosmic acceleration. We also have shown the determination of effective Newtonian constant $G_{eff}$ and how it satisfies the BBN and solar constraints.

We have summarised our results from the numerical analysis in Table (\ref{table:2})  marginalising the parameters by means of \texttt{Cobaya} sampler and \texttt{GetDist} package. In Fig.(\ref{fig:fig1}), we have presented the triangular plots with contours of the parameters ($\Omega_m, S_8, H_0$) and show a reduction of positive correlation between $S_8$ and $H_0$. Our best results have come from the joint dataset P18+R20 and indicates that the $H_0$ tension persists around $\sim 2.67\sigma$ and the $S_8$ tension is alleviated with tension below $1\sigma$ when compared with DESY1 in all dataset considered. We compared our results with KIDS+VIKING-450 and DESY1 combined and we obtained a tension around $\sim 1.34\sigma$ up to $\sim 1.5\sigma$. When compared with Kids1000 probe, we obtained similar results. From a larger dataset, with P18+\\R20+BAO+DESY1+Pantheon, our model also presented a ``promising'' performance (as proposed in Ref.(\cite{divalentino}), table B2) with the $H_0$ tension below $3\sigma$ at $68.4\%$ CL and below $2\sigma$ at $95.7\%$ CL, since the tensions between Planck and other probes sit well above three standard deviations. In all cases examined in $\beta$-model, the fluid parameter $w$ is preferred to be like that of a phantom dark energy in accordance with Planck data $w=-1.58^{+0.52}_{-0.41}$. We have noticed that the inclusion of BAO has slightly aggravated the tensions. Interestingly, this phantom behaviour corroborates ref. \cite{konitopoulos} that proposes a scalar-tensor Finsler cosmological model.

These results pose an interesting scenario since the model seems to provide a necessary gravitational strength to alleviate both $H_0$ and $S_8$ tensions for further analysis. This was obtained by the inclusion of the extrinsic curvature as a pivot element to modify standard Einstein's gravity. As prospects, we intend to analyse a combination with BAO+BBN on the Hubble tension with two free parameters of the model. The integrated Sachs–Wolfe (ISW) effect will be analysed in the light of larger surveys on dark energy and to study the impact of this model on CMB power spectrum with anisotropic parameters.

\section{Acknowledgements}
M.D. Maia for criticisms and suggestions. A. J. S. Capistrano thanks Funda\c{c}\~{a}o Arauc\'{a}ria/PR for the Grant CP15/2017-P$\&$D 67/2019. \\

\end{document}